\documentclass[a4paper,11pt]{article}
\pdfoutput=1

\usepackage{jheppub}

\makeatletter
\renewcommand\@fpheader{} 
\makeatother

\usepackage[utf8]{inputenc}
\usepackage[T1]{fontenc}
\usepackage[english]{babel}

\usepackage{comment}
\usepackage{url}
\usepackage{graphicx}
\usepackage{dcolumn}
\usepackage{bm}
\usepackage{amsmath}
\usepackage{amssymb}
\usepackage{subcaption}
\usepackage{slashed}
\usepackage{accents}
\usepackage{float}
\usepackage{soul}
\usepackage[normalem]{ulem}

\graphicspath{{images/}}

\newcommand{\be}{\begin{equation}}
\newcommand{\ee}{\end{equation}}
\newcommand{\bea}{\begin{eqnarray}}
\newcommand{\eea}{\end{eqnarray}}

\newcommand{\om}{\omega}
\newcommand{\eps}{\varepsilon}

\def\d{d\kern-.8 ex\vrule height 1.3 ex depth-1.24 ex width .7 ex \kern .15 ex}
\def\D{D\kern-1.7 ex\vrule height .87 ex depth-.8 ex width .7 ex \kern .95 ex}

\usepackage{mathtools}

\usepackage{xcolor}

\definecolor{darkcyan}{rgb}{0.0, 0.27, 0.67}

\definecolor{levcolor}{rgb}{0.67, 0.41, 0.69}

\definecolor{brown}{rgb}{0.56,0.22,0.22}

\title{Another Brick in the Wall: Spectral Form Factor of Normal Modes in Horizonless Geometries}

\author[a]{A.\,I.~Belokon,}
\author[b]{M.~\v{C}ubrovi\'c,}
\author[b]{V.~\DJ uki\'c,}
\author[c,d]{V.\,V.~Pushkarev,}
\author[e]{L.~Umrikhin}

\affiliation[a]{Department of Computational Mathematics and Mathematical Physics, Faculty of \mbox{Fundamental} Science, Bauman Moscow State Technical University, 2/18 Rubtsovskaya nab., \mbox{105005 Moscow}, Russian Federation}
\affiliation[b]{Center for the Study of Complex Systems, Institute of Physics Belgrade, University of Belgrade, Pregrevica 118, 11080~Belgrade, Republic of Serbia}
\affiliation[c]{National University of Science and Technology MISIS, Leninsky Prospekt 4, 119049~Moscow, Russian Federation}
\affiliation[d]{Steklov Mathematical Institute of Russian Academy of Sciences, 8 Gubkina St., 119333~Moscow, Russian Federation}
\affiliation[e]{Faculty of Physics, King's College London, WC2R 2LS~London, United Kingdom}

\emailAdd{
\\
belokon.ai14@physics.msu.ru
\\
cubrovic@ipb.ac.rs
\\
djukic@ipb.ac.rs
\\
pushkarev.vv@misis.ru 
\\
lev.umrikhin@kcl.ac.uk}

\abstract{
We compute the normal mode spectrum and its spectral form factor for a probe scalar wave in smooth horizonless backgrounds. We first consider the BTZ black hole with a brickwall that hides the near-horizon region, which has recently been proposed as a toy model for a finite-temperature semiclassical black hole microstate. In this setup, the spectral form factor may or may not have a linear ramp, depending on the details of the brickwall construction. However, we argue that even in cases that do show a ramp, this ramp does not originate from level correlations associated with quantum chaos. Then we calculate the normal mode spectrum for a genuine microstate geometry (superstratum) and find evidence of chaos, both with a ramp and level repulsion. We conclude that the eigenvalue spectrum of normal modes can indeed serve as a tool for investigating the quantum dynamics of fuzzballs; however, the brickwall models are incapable of capturing this phenomenon.}

\begin{document} 
\maketitle
\flushbottom

\section{Introduction and summary}
The relation between black holes and quantum chaos is well established in holography: concepts such as fast scrambling \cite{FastScramblers}, maximum chaos of the dual holographic CFT \cite{MSSbound}, and the structure of the energy spectrum and spectral form factor (SFF) consistent with the random matrix theory (RMT) universality \cite{Cotler:2016fpe} can almost be taken as the defining properties of black holes. However, there is an elephant in the room. The RMT description of quantum chaotic systems is about refining the discreteness of the spectrum of the quantum mechanical Hamiltonian. In other words, the RMT paradigm is about the deep quantum regime rather than the semiclassical chaos. On the other hand, holographic indicators of chaos such as the MSS bound~\cite{MSSbound} hold in the strongly coupled regime \emph{of the dual field theory}, meaning that on the gravity side we look at the classical (super)gravity regime. The natural question, therefore, goes as follows: how can we observe the manifestation of chaos directly in the bulk?

An immediate idea is to consider the leading quantum corrections to the bulk classical action. To date, this has been achieved only for low-temperature black holes close to extremality, both with \cite{Heydeman:2020hhw} and without \cite{Iliesiu:2020qvm} supersymmetry. The result is the celebrated Schwarzian sector that introduces corrections to the black hole entropy and free energy, at the same time pinpointing the structure of the spectrum which is closely related to SYK model \cite{Maldacena:2016hyu}. The geometric picture is that of a long but finite near-horizon AdS$_2$ sector where the Schwarzian excitations propagate \cite{Maldacena:2016upp}. Another idea is to move beyond the solutions of general relativity and consider instead explicit black hole microstates, in particular stringy constructions known as fuzzballs. These geometries look like black holes from far away while being at the same time completely smooth and horizonless. According to the fuzzball program, these geometries are conjectured to be individual microstates of a black hole, where the latter is a coarse-grained macroscopic description of an ensemble of such states~\cite{Bena:2022rna, Bena:2022ldq}.

Compared to the exact extremal black holes with their infinite AdS$_2$ throats, in near-extremal black holes and fuzzballs we have a finite AdS$_2$ region with a cap. It is not known yet whether we can derive an analogue of the Schwarzian sector in these cases but, given that both exhibit a capped AdS$_2$ region, it would be natural to expect that a similar formalism should apply, because both systems have a smooth limit to the infinite-throat regime. Given the crucial role of quantum chaos in determining the key features of black holes, it is of prime importance to understand if and how the bread-and-butter chaotic properties like fast scrambling and the presence of the RMT-like linear ramp in the SFF change upon capping off the extremal throat.

The fuzzball program also has problems summarized, e.g., in \cite{Raju:2018xue}. Our hope is to shed light on the role of fuzzballs by studying their dynamical and chaotic properties in detail, in order to compare them to the state-of-the-art knowledge of the black hole chaos. Part of the problem is the extremely complicated nature of fuzzball solutions, which makes them difficult to investigate analytically. In order to overcome this issue, a number of recent papers~\cite{Das:2022evy, Das:2023ulz, Das:2023yfj,  Das:2023xjr, Krishnan:2023jqn, Das:2024fwg, Banerjee:2024ivh, Jeong:2024jjn, Ageev:2024gem, Ageev:2025efa, Jeong:2025jyx, Caceres:2025qlh, Basu:2025zkp, Begines:2026fnx} have studied BTZ- and Schwarzschild-AdS black holes (and also several other spacetimes with horizons) with a hard wall cutoff in the near-horizon region in the spirit of the famous paper by 't Hooft \cite{tHooft:1984kcu}. This toy model, called in the aforementioned papers a synthetic fuzzball, is declared to mimic (in a crude way) the smooth cap of fuzzballs and near-extremal black holes. The idea is to cut off a part of the spacetime containing the horizon and its interior and impose the Dirichlet boundary condition for the bulk fields on the wall, as opposed to the infalling condition on the horizon in full black hole geometries. 

The papers on synthetic fuzzballs have found that some statistical indicators of the normal mode spectrum of a classical scalar field on such backgrounds resemble characteristic signatures usually attributed to quantum chaos in the context of RMT. Here is a short list of the main findings.
\begin{itemize}
\item \emph{Discrete normal mode spectrum.} Replacing an infalling boundary condition at the horizon by a reflecting condition at the wall turns the complex quasinormal mode problem into a normal mode problem with a real and discrete spectrum. The spectrum strongly depends on the radial overtone $n$ and has a much weaker dependence on the angular quantum number $J$ \cite{Das:2022evy,Krishnan:2023jqn}.


\item \emph{A dip-ramp-plateau structure without conventional level repulsion.} The SFF constructed from the partition sum over the normal frequencies $\omega_{n,J}$ of the form\footnote{In practice, one clearly has to impose finite cutoffs on $n$ and $J$.}:
\be
    Z(\beta+it)=\sum_{n,J}e^{-(\beta+it)\omega_{n,J}},
\ee
has been found to exhibit an apparent dip-ramp-plateau structure, with an approximately linear ramp, both with and without ensemble averaging. At the same time, the corresponding nearest-neighbor spacing distribution is not of the Wigner-Dyson form \cite{Das:2024fwg, Banerjee:2024ivh}. By restricting the spectral sum to different ranges of \(n\) and \(J\), the ramp was traced specifically to the weak and non-linear dependence of the normal mode frequencies on the angular quantum number~$J$, rather than to an approximately equidistant radial tower over $n$ \cite{Das:2022evy}.


\item \emph{Persistence across black hole geometries.} In rotating BTZ brickwall models, the grand canonical SFF was found to retain its dip-ramp-plateau form arbitrarily close to extremality. Exactly at extremality the nontrivial ramp disappears as the relevant spectrum becomes approximately linear. A ramp-like feature in the SFF of normal modes has also been reported for higher-dimensional brickwall AdS-Schwarzschild black holes and for brickwall models associated with cosmological horizons \cite{Das:2023xjr,Das:2024fwg,Ageev:2024gem,Begines:2026fnx}.

\end{itemize}

The fact that the SFF of the normal mode spectrum possesses a linear ramp naturally begs the question if maybe the brickwall model captures the quantum chaos of black holes at the (semi)classical\footnote{We write the word ``semi-'' in brackets because nothing is really quantized in this setup: we consider the SFF of the normal mode spectrum of a classical scalar field on a fixed classical background. The relation of this model to the underlying quantum features of black hole microstates is conjectured by the initiators of this program from the presence of the unit-slope linear ramp in the SFF, so that semiclassical effects may, in some way, be implicitly embedded in this setup.} level. On the other hand, one would intuitively think that by considering a (semi)classical problem on top of a spacetime with no horizon and interior, we do not probe the black hole quantum Hamiltonian or the set of its eigenstates in any way. The latter issue also holds for the normal modes of probes on top of genuine fuzzballs: does the statistics of normal frequencies tell us anything about the fuzzball itself?

Let us also stress that the reported combination of spectral signatures is highly unusual independently of any black hole interpretation. The homogeneous brickwall model is declared to combine an integrable classical problem, a non-Wigner-Dyson spacing distribution, a linear ramp in the SFF without disorder averaging and a nontrivial Krylov complexity peak. To the best of our knowledge, the closest extensively studied precedent for such a mismatch between standard diagnostics is arithmetic quantum chaos. Arithmetic hyperbolic billiards have strongly chaotic classical geodesic flows, yet their hidden Hecke symmetries and the exponentially large degeneracy of their periodic-orbit length spectra lead to nearly Poissonian local statistics and anomalous (generally non-RMT) form factors~\cite{Bogomolny:1992cj,Bogomolny:2003}. In the case under consideration, the underlying mechanisms need not be the same: in particular, the circular brickwall setup is classically integrable and has no known arithmetic symmetry. The comparison is nevertheless instructive because it makes clear that classical chaos, nearest-neighbor repulsion, long-range spectral rigidity, the slope of the SFF ramp and Krylov growth are logically independent diagnostics.

\subsection*{Summary}

In this paper, we analyze and compare the normal mode spectrum for brickwall models and for fuzzballs, specifically the superstratum \cite{Bena:2015bea,Shigemori:2020yuo}. Our goal is to understand where the linear ramp in brickwall models comes from, to see if it also exists for the normal modes in superstratum geometries, and whether it can be related to some well-defined quantum chaotic system. We conclude that it cannot: the linear ramp in brickwall backgrounds is only obtained for special, highly symmetric brickwall geometries, with a highly peculiar choice of quantum numbers, and disappears in more general brickwall setups. We are not in any disagreement with the previous works that find the ramp: we find it too, but conclude, from further analysis of the SFF, that it cannot be related to level correlations.

On the other hand, the normal mode spectrum in superstratum backgrounds shows a clear correlation with the outcomes of other, long-established criteria of quantum and wave chaos, such as the Berry random wave statistics for probe solutions. The single-mode superstratum~\cite{Bena:2015bea} shows no ramp, while the momentum-fractionated superstratum \cite{Bena:2016agb} has a ramp. We perform further study of its normal modes and confirm that they are indeed rooted in level correlations, the main sign of chaos. This is in line with the previous findings of~\cite{Djukic:2026ejk,Djukic:2026phb} that the scalar wave equation in the background of single-mode superstrata is only slightly nonseparable, and the corresponding wave chaos saturates at some point, while the momentum-fractionated superstrata yield strongly nonseparable dynamics with very strong wave chaos. 

Still, the connection between normal modes and chaos is not completely clear, even for superstrata. This is because the wave equation can only be packaged into a \emph{zero-energy} effective Klein-Gordon or Schr\"odinger equation, where the physical energy, i.e., the normal mode frequency $\omega$, only enters as a parameter, typically in a nonlinear way. On a physical level, this simply means that the eigenlevels of a \emph{probe} do not necessarily follow the (presumably chaotic) dynamics of the eigenlevels of the background. On top of that, fuzzballs are typically BPS states, and the supersymmetry imposes a high degeneracy of the spectrum; the spectrum of a non-supersymmetric probe, thus, clearly cannot have much in common with the spectrum of the background\footnote{\label{foot:microstr}Smooth non-BPS microstates, called microstrata, are also known, see \cite{Ganchev:2021pgs, Ganchev:2023sth}. They are, however, special nonextremal solitonic states and do not yet provide a comparably controlled description of a typical finite-temperature black hole microstate.}. The fact that we nevertheless detect the ramp as a sign of chaos in normal modes is likely related to the fact that the same equation governs the \emph{eigenstates} of the wave operator, which were argued in \cite{Djukic:2026ejk,Djukic:2026phb} to detect the chaoticity of the microstate from a reasoning akin
to the eigenstate thermalization hypothesis (ETH): the RMT-like structure of the
off-diagonal elements of a simple operator stems precisely from the properties of the
eigenstate (since the operator is simple, and in addition, in a BPS background, its
spectrum will necessarily be degenerate).

One might now be disappointed in the brickwall model and conclude that it not only fails to capture the black hole dynamics but also does not mimic the behavior of smooth microstate solutions on the level of a toy model. Nevertheless, it seems to us that the brickwall model can still be interesting in some other respects: 
\begin{itemize}
    \item It might still be an analytically tractable proxy for a finite-temperature fuzzball cap. As we have mentioned, known microstate geometries are overwhelmingly BPS, and therefore extremal (see the footnote \ref{foot:microstr} for the rare exceptions).
    \item The normal mode spectrum of this model resembles the discreteness of the quantum gravity spectrum of AdS black holes. 
\end{itemize}

The plan of the paper is the following. In Section~\ref{sec:setup}, we define the brickwall setup, namely, the geometry and the properties of the normal mode spectrum of the massless Klein-Gordon equation in this background. In Section~\ref{sec:circbrick}, we analyze the corresponding SFF of this spectrum, confirm the existence of the ramp in one case, and show that the ramp comes from a specific scaling of the spectrum, not from level correlations and chaos. In Section~\ref{sec:angdepwall}, we consider a deformed (elliptical) shape of the wall and show that the level statistics do not display a linear ramp at all. In Section~\ref{sec:ss}, we compute the normal modes for a superstratum, find the ramp for the momentum-fractionated case, and discuss the implications of our findings for the brickwall model. Section~\ref{sec:conclusions} sums up our conclusions.



\section{Setup}
\label{sec:setup}

\subsection{Normal modes of BTZ black hole with a brickwall}
\label{subsec:brickwall-btz}
\emph{Circular wall.} We work in three-dimensional Einstein gravity with negative
cosmological constant:
\begin{equation}
    I_{\mathrm{grav}}
    =
    \frac{1}{16\pi G_3}
    \int\limits_{\mathcal M} d^3x\,\sqrt{-g}
    \left(
        R+\frac{2}{\ell^2}
    \right)
    +I_{\mathrm{bdy}},
    \label{eq:gravity-action}
\end{equation}
where $\ell$ is the AdS radius and $I_{\mathrm{bdy}}$ denotes the corresponding boundary terms appropriate to the variational problem. The static BTZ black hole has the following metric:
\begin{equation}
    ds^2
    =
    -f(r)\,dt^2
    +\frac{dr^2}{f(r)}
    +r^2d\phi^2,
    \qquad
    f(r)
    =
    \frac{r^2-r_h^2}{\ell^2},
    \qquad
    \phi\sim\phi+2\pi.
    \label{eq:static-btz-metric}
\end{equation}
Here, $r_h$ is the radial coordinate of the horizon.

Following the original 't Hooft prescription~\cite{tHooft:1984kcu} and many subsequent works~\cite{Das:2022evy,Krishnan:2023jqn,Ageev:2024gem,Das:2024fwg,Gayari:2026jwn}, we cut off the region $r<r_w$, which includes the interior $r < r_h$, and restrict the geometry to the following manifold:
\begin{equation}
    \mathcal M_w
    =
    \left\{
        (t,r,\phi)
        \,\middle|\,
        r_w\leq r<\infty
    \right\},
    \qquad
    r_w=r_h+\varepsilon,
    \qquad
    \varepsilon>0.
    \label{eq:brickwall-domain}
\end{equation}
In the context of the synthetic fuzzball research program~\cite{Das:2022evy,Das:2023ulz,Das:2023xjr,Krishnan:2023jqn,Das:2024fwg,Banerjee:2024ivh,Jeong:2024jjn,Ageev:2024gem,Basu:2025zkp,Ageev:2025efa,Jeong:2025jyx,Caceres:2025qlh,Begines:2026fnx}, the idea is to emulate a black-hole-like smooth horizonless geometry, i.e. a finite-temperature microstate geometry, in the simplest possible way, in order to distill the consequences of horizonlessness as such. The position of the wall $\varepsilon$ is not vanishing: it is a finite constant. The timelike inner boundary $r=r_w$ is, therefore, perfectly reflecting (the so-called ``brickwall''). 


As a probe of dynamics, we consider a minimally coupled massless scalar field on the background~$\mathcal{M}_w$~\eqref{eq:brickwall-domain}. The action and the corresponding Klein-Gordon equation of motion read:
\begin{equation}
    I_\Phi
    =
    -\frac{1}{2}
    \int\limits_{\mathcal M_w}
    d^3x\,\sqrt{-g}
        g^{\mu\nu}
        \partial_\mu\Phi\,
        \partial_\nu\Phi,
    \label{eq:scalar-action-KG}
\end{equation}
\begin{equation}
    \Box\Phi=0 \quad + \quad \text{boundary conditions on $\partial\mathcal{M}_w$}.
    \label{eq:klein-gordon}
\end{equation}
Since the background~\eqref{eq:brickwall-domain} is invariant under the translations of $t$ and $\phi$, we use the ansatz:
\begin{equation}
    \Phi(t,r,\phi)
    =
    e^{-i\omega t+iJ\phi}
    R_{\omega J}(r),
    \qquad
    J\in\mathbb Z,
    \label{eq:mode-ansatz}
\end{equation}
where $\omega$ and $J$ are the energy and the angular momentum, respectively. The equation for the radial part $R_{\omega J}(r)$ of the ansatz~\eqref{eq:mode-ansatz} reads:
\begin{equation}
    \frac{1}{r}
    \frac{d}{dr}
    \left(
        rf(r)
        \frac{dR_{\omega J}}{dr}
    \right)
    +
    \left(
        \frac{\omega^2}{f(r)}
        -\frac{J^2}{r^2}
    \right)
    R_{\omega J}(r)
    =0.
    \label{eq:radial-equation-general}
\end{equation}
After imposing the normalizability condition of the solution to~\eqref{eq:radial-equation-general} at the AdS boundary $r \to \infty$, its near-boundary asymptotic takes the form:
\begin{equation}
R_{\omega J}(r)\sim B_{\omega J}r^{-\Delta_+}\qquad\Delta_+=1+\nu,\qquad\nu=\sqrt{1+\mu^2\ell^2}.
\label{eq:ads-falloffs}
\end{equation}
At the brickwall $r = r_w$, we impose the Dirichlet condition:
\begin{equation}
    R_{\omega J}(r_w)=0,
    \label{eq:wall-dirichlet}
\end{equation}
which defines the discrete normal mode spectrum: \begin{equation}
    \{\omega_{n,J}\},
    \qquad
    n\in\mathbb N,
    \qquad
    J\in\mathbb Z.
    \label{eq:brickwall spectrum}
\end{equation}
Since the radial equation~\eqref{eq:radial-equation-general} depends on the angular momentum only through $J^2$, the spectrum~\eqref{eq:brickwall spectrum} is parity-invariant under $J\longrightarrow-J$. The derived spectrum~$\omega_{n,J}$ is the normal-mode analogue of the quasinormal-mode spectrum of black holes~\cite{Ichinose:1994rg, Keski-Vakkuri:1998gmz}, except that, in the absence of horizon and dissipation, the frequencies are real (in other words, the Klein-Gordon operator is self-adjoint).

\emph{Angle-dependent wall.} The circular wall $r = r_w$ preserves the Killing symmetry generated by $\partial_\phi$, which is also present for the corresponding static non-rotating BTZ black hole. One may therefore expect that the dynamical signatures of these two backgrounds are not too different. In order to test the importance of this Killing symmetry, we will also consider an angle-dependent (elliptical) brickwall, which explicitly breaks the circular symmetry. Time-translation invariance still remains, so the appropriate ansatz takes the form:
\begin{equation}
\Phi(t,r,\phi)=e^{-i\omega t}\psi(r,\phi).\label{eq:phideformed}
\end{equation}
In this case, the angular momentum is no longer conserved, and different angular sectors mix. 

\subsection{\label{subsec:spectral-data}BTZ normal mode spectrum}

The eigenvalue problem~\eqref{eq:radial-equation-general}--\eqref{eq:wall-dirichlet} defines a two-parameter set of positive real frequencies~\eqref{eq:brickwall spectrum}~\cite{Ichinose:1994rg}. Since the spectrum~\eqref{eq:brickwall spectrum} is labeled by two quantum numbers, the notions of nearest neighbor and nearest-neighbor spacings are not well defined until a one-dimensional spectral sequence $\mathcal S$ has been selected. Two natural choices are the \emph{fixed-overtone spectrum} or \emph{angular spectrum} $\mathcal{S}_n(J)$ and the \emph{fixed-angular-momentum spectrum} or \emph{radial spectrum}~$\mathcal{S}_J(n)$:
\begin{eqnarray}
\mathcal{S}_n(J)=\left\{\omega_n(J)\right\}_{J\in\mathbb{Z}}\equiv\left\{\omega_{n,J}\right\}_{n=\mathrm{const},J\in\mathbb{Z}},\nonumber\\
\mathcal S_J(n)=\left\{\omega_J(n)\right\}_{n\in\mathbb{N}}\equiv\left\{\omega_{n,J}\right\}_{J=\mathrm{const},n\in\mathbb{N}}.
\label{eq:spectral-subsequences}
\end{eqnarray}
Previous analyses of the BTZ brickwall normal mode spectra (first performed in \cite{Das:2022evy}) were focused primarily on $\mathcal S_{n}(J)$, i.e., on the angular spectrum at fixed radial overtone~$n$. We will discuss below the caveats with this sequence.


Once a one-dimensional spectral sequence has been chosen, we arrange its frequencies in non-decreasing order and relabel them by a single ordering index:
\begin{equation}
    \mathcal S
    =
    \left\{
        \omega_a
    \right\}_{a=1}^{N_{\mathcal S}},
    \qquad
    \omega_1\leq\omega_2\leq\ldots\leq\omega_{N_{\mathcal S}},
    \qquad
    N_{\mathcal S}\equiv|\mathcal S|.
    \label{eq:ordered-spectral-sample}
\end{equation}
These positive normal mode frequencies are then regarded as the energy levels entering the SFF. The spectral density of the sequence $\mathcal S$ is defined in the usual way:
\begin{equation}
    \rho_{\mathcal S}(\omega)
    =
    \sum_{a=1}^{N_{\mathcal S}}
    \delta\left(
        \omega-\omega_a
    \right),
    \label{eq:empirical-density}
\end{equation}
so the analytically continued spectral partition function reads:
\begin{equation}
    Z_{\mathcal S}(\beta+it)
    =
    \sum_{a=1}^{N_{\mathcal S}}
    e^{-(\beta+it)\omega_a}.
    \label{eq:continued-partition-sum}
\end{equation}
With \eqref{eq:continued-partition-sum} we can define the non-normalized SFF $K_\mathcal{S}(\beta,t)$ and the normalized SFF $g_\mathcal{S}(\beta,t)$ as follows:
\begin{eqnarray}
K_{\mathcal S}(\beta,t)&=&\left|Z_{\mathcal S}(\beta+it)\right|^2 = \sum_{a,b=1}^{N_{\mathcal S}}e^{-\beta(\omega_a+\omega_b)}e^{-it(\omega_a-\omega_b)}\label{eq:unnormalized-sff}\\
g_{\mathcal S}(\beta,t)&=&\frac{K_{\mathcal S}(\beta,t)}{K_{\mathcal S}(\beta,0)}.\label{eq:normalized-sff}
\end{eqnarray}
Using the spectral density \eqref{eq:empirical-density}, non-normalized SFF \eqref{eq:unnormalized-sff} can equivalently be written as:
\begin{align}
    K_{\mathcal S}(\beta,t)
    &=
    \int\limits_0^\infty d\omega_1
    \int\limits_0^\infty d\omega_2\,
    e^{-\beta(\omega_1+\omega_2)}
    e^{-it(\omega_1-\omega_2)}
    \rho_{\mathcal S}(\omega_1)
    \rho_{\mathcal S}(\omega_2).
    \label{eq:sff-density-representation}
\end{align}
In order to smooth out the strong short-time fluctuations of the raw SFF~\eqref{eq:unnormalized-sff} and to better appreciate the overall trends, we perform a moving average over a time window $\Delta t$\footnote{In some situations, the averaging of the SFF has deeper meaning. Famously, when considering the SFF of dynamical JT gravity or dynamical AdS${}_3$, it should likely be interpreted as a genuine average over couplings or theories, at least at some mesoscopic level. But we work on a fixed, non-dynamical background, and our partition function and the SFF include only the dynamics of the probe field. This means that, in fact, there is nothing to average over at the fundamental level, and our running average is merely a convenient computational tool.}:
\begin{equation}
K_{\mathcal S}^{(\Delta t)}(\beta,t)\equiv\int dt'\,W_{\Delta t}(t-t')K_{\mathcal S}(\beta,t'),
\qquad\int dt'\,W_{\Delta t}(t')=1.
\end{equation}

\subsection{WKB approximation for BTZ normal mode spectrum}

To estimate the energies of the spectra $\mathcal S_{n}(J)$ and $\mathcal S_{J}(n)$, let us consider the WKB approximation for scalar waves on the brickwall background~\eqref{eq:brickwall-domain}. As it was shown in \cite{Das:2023xjr}, the WKB condition for a scalar wave in the background of a static BTZ black hole with a circular brickwall results in the following spectrum:
\be
    \om_{n, J} \simeq \frac{A_n}{W \displaystyle\left(\sqrt{\frac{2}{\eps}}\frac{A_n}{J}\right)}, \qquad A_n = \frac{\pi(8n + 3)}{4}, \qquad n \in \mathbb{Z}_+, \quad J \in \mathbb{Z},
    \label{eq:om_WKB}
\ee
where $W(z)$ is the Lambert $W$-function. This approximation is valid within the range\footnote{Although, in general, we consider various wall positions, not necessarily at a small distance from the horizon, for the WKB approximation to work, we do need small $\varepsilon$.}:
\be
    1 \ll \om_{n, J} \ll |J|, \qquad 1 \ll |J| \lesssim \eps^{-2/3}, \qquad \eps \to 0.
    \label{eq:WKB-validity}
\ee
For the radial spectrum $\mathcal{S}_J(n)$, the angular momentum $J$ is fixed. At large radial quantum number $n$, we get $A_n \simeq 2\pi n$, and therefore:
\begin{equation}
    \omega_{J}(n)
    \simeq
    \frac{4\pi n}
    {\log\!\left(\dfrac{4\pi n}{\varepsilon J}\right)},
    \qquad n \gg 1,
    \label{eq:radial-WKB-asymptotics}
\end{equation}
where $n$ is kept inside the WKB validity window. The last expression retains only the leading large-argument asymptotic of the Lambert $W$-function. Thus, the radial spectrum is approximately linear in $n$, up to a slowly-varying logarithmic correction.

For the angular spectrum $\mathcal{S}_n(J)$, on the other hand, the radial overtone~$n$ is fixed. The dependence of $\mathcal{S}_n(J)$ on $J$ can be reduced to a logarithmic expansion in the vicinity of some $J_0$ chosen within the window~\eqref{eq:WKB-validity}\footnote{More precisely, the expansion is valid for $2|\alpha_n|/|\gamma_n|\gg\left|\log\frac{J}{J_0}\right|$.} \cite{Das:2022evy}:
\be
    \om_n(J) = \om_* + \alpha_n \log\frac{J}{J_0} + \frac{\gamma_n}{2}\log^2\frac{J}{J_0} + \ldots,\label{eq:wkblog}
\ee
where:
\be
    \omega_* = \frac{A_n}{w_0}, \quad \alpha_n = \frac{A_n}{w_0(1+w_0)}, \quad \gamma_n = \frac{A_n(1+2w_0)}{w_0(1+w_0)^3}, \quad w_0 = W \displaystyle\left(\sqrt{\frac{2}{\eps}}\frac{A_n}{J_0}\right).
\ee
Therefore, in the leading approximation, the angular spectrum is logarithmically dependent on the appropriate quantum number $J$.

\section{Circular brickwall}
\label{sec:circbrick}
\subsection{Classical circular brickwall billiard}

We start with a simple exercise giving us intuition about the radial spectrum and the absence of the ramp or chaos in it. The idea is that the classical probe field on the BTZ background with a circular brickwall becomes equivalent to a classical billiard in the eikonal limit. The null geodesic Hamiltonian for the BTZ metric~\eqref{eq:static-btz-metric} is given by:
\be
    H
    =
    \frac12 g^{\mu\nu}p_\mu p_\nu
    =
    \frac12
    \left[
        -\frac{p_t^2}{f(r)}
        +f(r)p_r^2
        +\frac{p_\phi^2}{r^2}
    \right] = 0.
    \label{eq:btz-geodesic-hamiltonian}
\ee
The Killing vectors $\partial_t$ and $\partial_\phi$ give two conserved quantities, $E=-p_t$ and $L=p_\phi$. Reflections on the wall act on the momentum according to the Snellius law:
\begin{equation}
    p_\mu^{\mathrm{out}}
    =
    p_\mu^{\mathrm{in}}
    -
    2
    \left(
    p_\nu^{\mathrm{in}}n^\nu
    \right)n_\mu,
    \label{eq:covariant-reflection-law}
\end{equation}
where $n^\mu$ is the outward-pointing unit normal to the wall, $r=r_w$, given by:
\be
    n^\mu
    =
    \sqrt{f(r_w)}\,\delta^\mu_r.
    \label{eq:circular-wall-normal}
\ee
Since the normal has only one (radial) component, this reduces to:
\begin{equation}
    p_r^{\mathrm{out}}
    =
    -p_r^{\mathrm{in}},
    \qquad
    p_t^{\mathrm{out}}
    =
    p_t^{\mathrm{in}},
    \qquad
    p_\phi^{\mathrm{out}}
    =
    p_\phi^{\mathrm{in}},
    \label{eq:circular-wall-reflection}
\end{equation}
and together with the null constraint~\eqref{eq:btz-geodesic-hamiltonian}, this means that the reflection on the brickwall preserves both $E$ and $L$.

We conclude that the corresponding Hamilton-Jacobi equation is separable and the classical dynamics is integrable. Therefore, according to the Berry---Tabor~\cite{BerryTabor1977} conjecture, the corresponding quantum-mechanical energy spectrum~\eqref{eq:brickwall spectrum} \emph{within one symmetry sector in $J$}, i.e., for the radial spectrum~$\mathcal{S}_J(n)$, is not expected to obey the Wigner-Dyson statistics, and the SFF of this spectrum is also not expected to show the dip-ramp-plateau structure.

\subsection{Analysis of spectral form factor}

In~\cite{Das:2022evy} and subsequent works, the angular spectrum was chosen as a phenomenological indicator of chaos in the SFF of BTZ normal modes. This choice was motivated by the fact that for the spectrum $\mathcal{S}_J(n)$ no ramp is expected since this spectral sequence is approximately linear, as we have mentioned in eq.~\eqref{eq:radial-WKB-asymptotics}. This was also found in numerical calculations~\cite{Das:2022evy}, confirming the absence of the ramp. The angular spectrum $\mathcal{S}_n(J)$ was found to be more interesting phenomenologically. However, we will now argue that this choice is not a proper energy spectrum, as it mixes different selection sectors, i.e., different angular momentum quantum numbers, which is generically not allowed in spectral analysis.

\subsubsection{Symmetry-sector mixing}

For the circular wall, the quantum number $J$ labels an
exact symmetry sector. Rotational invariance implies:
\begin{equation}
    \left[\hat H,\hat J\right]=0,
    \qquad
    \hat J=-i\partial_\phi ,
    \label{eq:HJ-commutator}
\end{equation}
hence, the Hilbert space $\mathcal H$ and Hamiltonian $\hat H$ decompose according to:
\begin{equation}
    \mathcal{H}
    =
    \bigoplus_{J\in\mathbb{Z}}\mathcal{H}_J,
    \qquad
    \hat H
    =
    \bigoplus_{J\in\mathbb{Z}} \hat H_J,
    \qquad
    \operatorname{Spec}(\hat H_J)
    =
    \left\{\omega_{n,J}\right\}_{J=\mathrm{const},n\in\mathbb{N}}.
    \label{eq:symmetry-decomposition}
\end{equation}
Of the two sequences introduced in \eqref{eq:spectral-subsequences}, the first one, $\mathcal S_J(n)$, is the spectrum of the symmetry-reduced operator $\hat H_{J}$ for some fixed $J$. This sequence is (in principle) the appropriate object for performing the proper symmetry-resolved spectral analysis usually done in the literature on quantum chaos~\cite{Haake1991}. 

\begin{figure}
    \centering
    \includegraphics[width=0.5\linewidth]{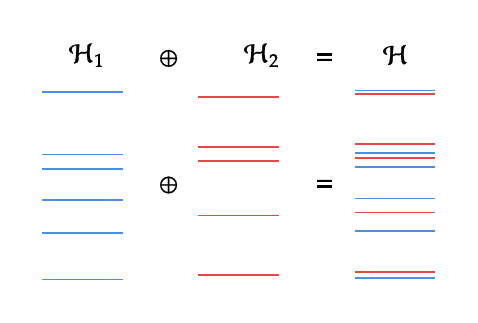}
    \caption{Level clustering of energy levels of a toy two-block Hamiltonian $H = H_1 \oplus H_2$, whose individual symmetry-reduced parts $\mathcal H_{1,2}$ have level-repulsive spectra. The same mechanism can, in principle, lead to a non-trivial statistics of the energy levels (or various dispersion branches, analogous to~\eqref{eq:transverse-sequence}) of the full Hamiltonian, ${H = \bigoplus\limits_{i} H_i}$, whose symmetry-reduced parts $H_i$ have level-repulsive spectra.}
    \label{fig:level-clustering}
\end{figure}

The second sequence, $\mathcal S_{n}(J)$, has a fundamentally different meaning. It contains one selected eigenvalue from each of different symmetry blocks:
\begin{equation}
    \mathcal S_{n}
    =
    \left\{
        \text{the $n$-th eigenvalue of }H_J
    \right\}_{J\in\mathbb{Z}}
    \subset
    \operatorname{Spec}(H).
    \label{eq:transverse-sequence}
\end{equation}
The radial label $n$ is just a label within every $J$-sector; it is not the eigenvalue of a symmetry generator whose eigenspace would define a natural invariant sector common to all $J$. Consequently, $\mathcal S_{n}(J)$ is neither the full spectrum of $\hat H$ nor the spectrum of any fixed closed symmetry block $\hat H_J$. It is well-known (e.g., \cite{Haake1991}), that testing the Wigner-Dyson statistics on level sequences that mix different symmetry sectors is risky, as the mixing can lead to level crossing or clustering even in a chaotic system, because rotational invariance forces some elements of the Hamiltonian reduced to the given $n$-sector from different $J$-sectors to be zero (see fig.~\ref{fig:level-clustering}).


The outcome of this reasoning is that, when considering $\mathcal{S}_n(J)$, a block-diagonal Hamiltonian is compared with a single random matrix, even though different blocks do not repel each other. Therefore, the apparent ramp in $\mathcal S_n(J)$ cannot diagnose spectral correlations within an irreducible dynamical sector, and should not be considered as evidence of quantum chaos in the sense of the Bohigas-Giannoni-Schmit conjecture.

\subsubsection{Unfolded angular spectrum}

We have explained that mixing different symmetry sectors in the angular spectrum makes its SFF questionable as an indicator of RMT level correlations, mainly because this mixing can give rise to fake level clustering, which might mask level repulsion present in individual symmetry-reduced parts. Now, however, we want to understand how to interpret the opposite situation: the presence of the linear ramp in the SFF of the normal mode angular spectrum, despite the fact that this spectral sequence is mixed. 

The level distribution in any spectrum is, generally speaking, shaped by two different ingredients: the smooth contribution described by the classical phase space density (which is also the leading semiclassical contribution to the density of states), and the oscillatory component stemming from level correlations. The universal RMT ramp arises from the second ingredient. When computing the level spacing statistics, it is necessary to perform the unfolding of the spectrum, i.e. to isolate the local correlations from the smooth envelope. For the SFF calculation (in the canonical ensemble, which is usually the case in high-energy context, and also in this work), this is not necessary; indeed, the statistical weights that enter the partition functions are sensitive to absolute values of the energies, not just fluctuations.

Our strategy is to unfold the spectrum and look at the unfolded SFF -- despite the fact that such SFF will not be physical anymore. If the ramp is still present, then it comes from level correlations; if not, then it comes from the smooth component. We reiterate that this is just a tool for investigating the origin of the ramp. The correct, physical SFF is the one without unfolding. This trick was already used in~\cite{Cotler:2016fpe} to determine the absence of RMT-type correlations between individual levels.


\emph{Logarithmic spectrum.} In \cite{Basu:2025zkp}, it is shown analytically that the leading-order WKB contribution to normal mode angular spectrum is of the logarithmic form:
\be
    \omega_n \sim \omega_* + \alpha\log n,
    \label{logspec}
\ee
and yields a linear ramp, which is conjectured to be a semiclassical manifestation of the hypothetical underlying RMT statistics in brickwall setups. While we are in perfect agreement with the existence of the linear ramp in this synthetic spectrum, we want to discuss in detail its microscopic origin in this particular construction. 

Consider first a spectrum that is completely specified by a smooth monotone function such as the one in (\ref{logspec}):
\be
\om_n = f(n),
\ee
with no additional fluctuations. The cumulative distribution function of a spectrum~$\{\omega_n\}$ is defined as the counting function of the number of levels lying below some value $\omega$:
\be
    \mathcal{N}(\om) \equiv \#\{n\,|\, \omega_n \leq \omega\} = \sum\limits_n\theta(\omega-\omega_n) =  f^{-1}(\om),
    \label{eq:staircase}
\ee
where $\theta(x)$ is the Heaviside step function. The unfolded levels are then defined as:
\be
    \widetilde{\om}_n = \mathcal{N}(\om_n) = f^{-1}\!\left(f(n)\right) = n.
\ee
Therefore, the unfolded spectrum $\{\widetilde{\omega}_n\}$ of an initial spectrum $\{\omega_n\}$ defined by some smooth trend~$f(n)$, as in~\eqref{logspec}, is exactly equidistant. Such a spectrum, akin to a linear harmonic oscillator, is evidently not indicative of chaos. The reason is that the presence of the linear ramp is a necessary -- not a sufficient -- condition for strong level correlations. In other words, the linear ramp of the exact logarithmic spectrum~\eqref{logspec} is a smooth-density effect rather than a consequence of correlations. It should therefore not be identified with the conventional random-matrix ramp. If the spectrum $\{\omega_n\}$ truly contained fluctuations $\delta\omega_n$ following random-matrix statistics, these would manifest as level fluctuations superimposed on this trend, $\omega_n = f(n) + \delta\omega_n$, and the unfolding procedure would simply make them clearly apparent on top of the equidistant spectrum:
\be
\widetilde{\omega}_n = f^{-1}(f(n)+\delta\omega_n) \simeq n + \frac{\delta\omega_n}{f'(n)}.
\label{eq:unfolded-seq-def}
\ee

\emph{Numerically computed spectrum.} The logarithmic dependence of the frequencies on $J$ is approximate; even the analytical WKB formula only yields a logarithmic dependence at leading order, and the exact values further deviate also from the WKB values. In order to check that the above unfolding reasoning remains valid, we have computed the SFF (physical and unfolded) with high-accuracy numerical results for the frequencies.

Unfolding changes mean density, hence, the time axis should be rescaled in order to compare the raw, i.e., physical SFF and the unfolded SFF. We measure time in units of the corresponding Heisenberg time $t_{\rm H}$ and denote the rescaled time by $\tau$:
\begin{equation}
    \tau = \frac{t}{t_{\rm H}},
    \qquad
    t_{\rm H} = 2\pi \bar\rho
    = \frac{2\pi}{\overline{\Delta\omega}},
\end{equation}
where $\bar\rho$ and $\overline{\Delta\omega}$ denote the mean spectral density and the mean level spacing, respectively, within the spectral window under consideration \cite{Haake1991}. Expressed in these units, the two SFF curves can be compared.

\begin{figure}[h!]\centering
    \subfloat[raw]{\includegraphics[width=0.48\linewidth]{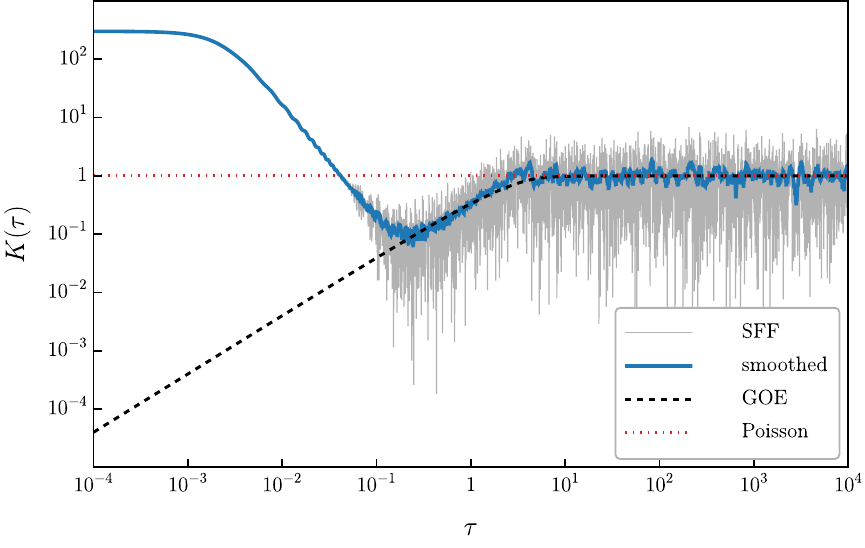}}\quad
    \subfloat[unfolded]{\includegraphics[width=0.48\columnwidth]{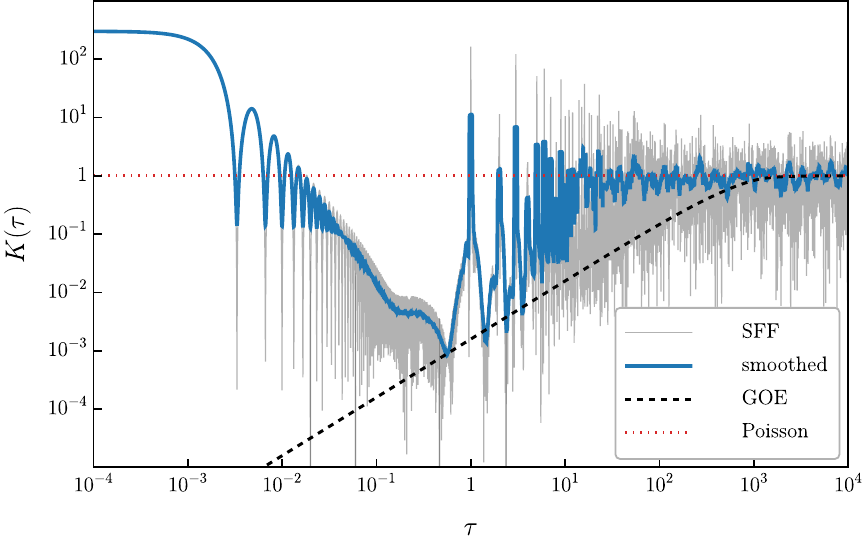}}
    \caption{The SFF $K(\tau)$ (Heisenberg time $\tau_H=1$) of the normal mode spectrum for a scalar field in BTZ black hole with a brickwall for (a)~raw, and (b)~unfolded angular spectrum $\mathcal{S}_n(J)$. The parameters are: ${n = 1}$, ${r_h = \ell = 1}$, and ${\beta = 0}$, and the sum in $K(\tau)$ is taken over 300 $J$-modes. Numerical unfolding of the angular spectrum is performed by a polynomial fit of degree~5. The blue curves show the moving averages of the gray ones. The dashed black line demonstrates the ramp part of the GOE ensemble, while the dotted red line denotes the Poisson plateau for reference. We conclude then that the ramp part of the brickwall BTZ normal mode SFF is driven by the smooth logarithmic trend~\eqref{logspec} and is not generated by universal level correlations.}
    \label{fig:unfolded_WKB}
\end{figure}

The comparison of the SFF with and without unfolding is given in fig.~\ref{fig:unfolded_WKB} and verifies our initial conclusions: the left-hand (physical) SFF is performed without unfolding and yields a clear linear ramp; the right-hand (unfolded) SFF performed with unfolding loses the ramp, which would not happen in a genuinely quantum chaotic system (in the RMT sense), since the unfolding should mainly influence the plateau, where the discrete level structure plays a role. The loss of the ramp in the formal SFF obtained after unfolding the spectrum is something that would not happen with a RMT ensemble.

\section{Angle-dependent brickwall}
\label{sec:angdepwall}





As we have shown in the previous section, the ramp of the raw SFF of the normal mode spectrum in the BTZ black hole with a circular brickwall does not indicate the RMT level correlations. Since this ramp is in fact caused by the logarithmic trend~\eqref{logspec} of the angular spectrum\footnote{In \cite{Das:2023ulz}, the robustness of the ramp was checked by introducing Gaussian random noise in the boundary conditions. In our opinion, this argument is not quite convincing: by reading off the statistics of the spectrum constructed this way we could well have ended up with the Gaussian statistics put in the system by hand.}, now the natural question is what happens with this ramp when the circular symmetry is broken and $J$ is not a good quantum number.

In this section, we construct a more general, non-circularly symmetric brickwall, in order to check if maybe genuine chaos will arise when the geometry does not have a high degree of symmetry. To that end, we consider the (simplest) angle-dependent brickwall, namely, an elliptical wall:
\be
    r_w(\phi)
    =
    \frac{ab}
    {\sqrt{a^2\sin^2\phi+b^2\cos^2\phi}},
    \label{eq:ell-wall}
\ee
where $a$ and $b$ are the major and minor semiaxes. In this case angular momentum is no longer conserved, and the SFF of normal modes can be calculated for other symmetry sectors, with no connection to $J$. 

The only remaining symmetries are discrete (two reflections with respect to major and minor symmetry axes). They generate four symmetry sectors and have no associated conserved quantities, so we do not expect the reflections to have a drastic effect on the spectrum. The Hilbert space $\mathcal{H}$ decomposes as:
\be
\mathcal{H} = \mathcal{H}_{++} \oplus \mathcal{H}_{+-} \oplus \mathcal{H}_{-+} \oplus \mathcal{H}_{--},
\label{eq:H-ellipse}
\ee
where the subscripts denote the signs of parity under each of the corresponding reflections. 

For the ansatz (\ref{eq:phideformed}), the Klein-Gordon equation in the background~\eqref{eq:brickwall-domain} with an elliptical wall~\eqref{eq:ell-wall} becomes:
\be
\frac{1}{r}\partial_r\left(rf\partial_r\psi\right)+\frac{1}{r^2}\partial_\phi^2\psi+\frac{\omega^2}{f}\psi=0,
\label{eq:KG-ellipse}
\ee
with the Dirichlet boundary condition $\psi(r_w(\phi), \phi)=0$.

\begin{figure}[!htbp]
    \centering
    \begin{minipage}[t]{0.42\textwidth}
        \centering
        raw\par\medskip

        \begin{subfigure}{\linewidth}
            \includegraphics[width=\linewidth]{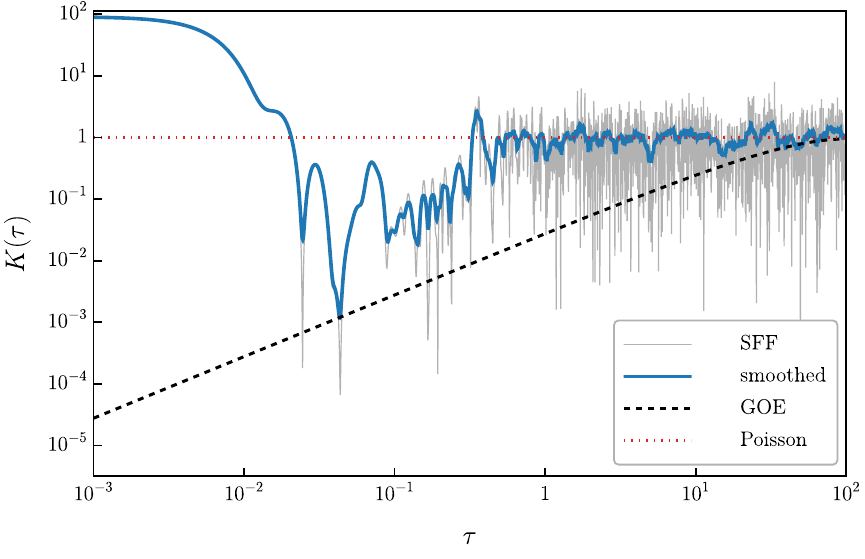}
            \caption{$(--)$}
        \end{subfigure}
        \par\smallskip

        \begin{subfigure}{\linewidth}
            \includegraphics[width=\linewidth]{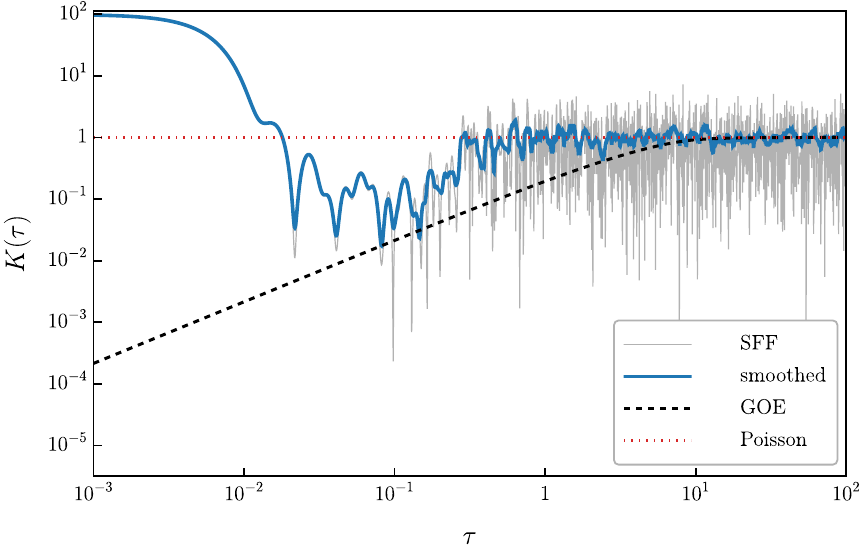}
            \caption{$(-+)$}
        \end{subfigure}
        \par\smallskip

        \begin{subfigure}{\linewidth}
            \includegraphics[width=\linewidth]{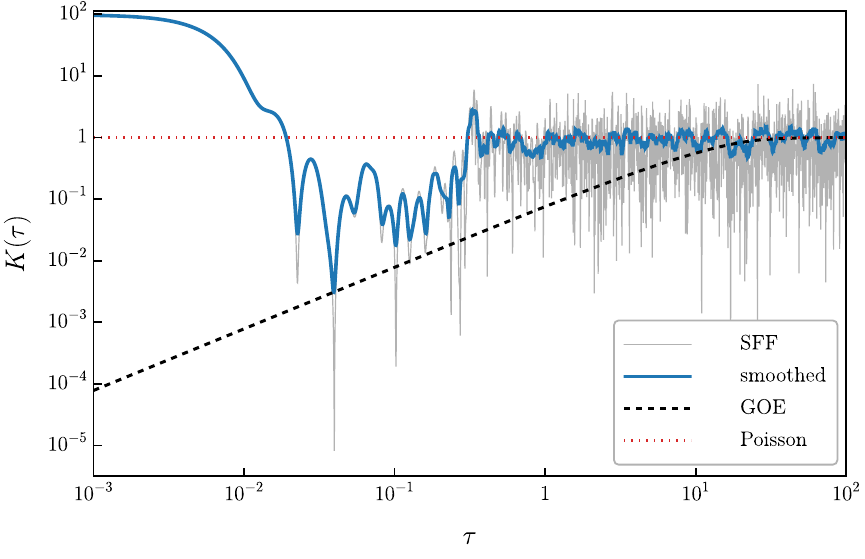}
            \caption{$(+-)$}
        \end{subfigure}
        \par\smallskip

        \begin{subfigure}{\linewidth}
            \includegraphics[width=\linewidth]{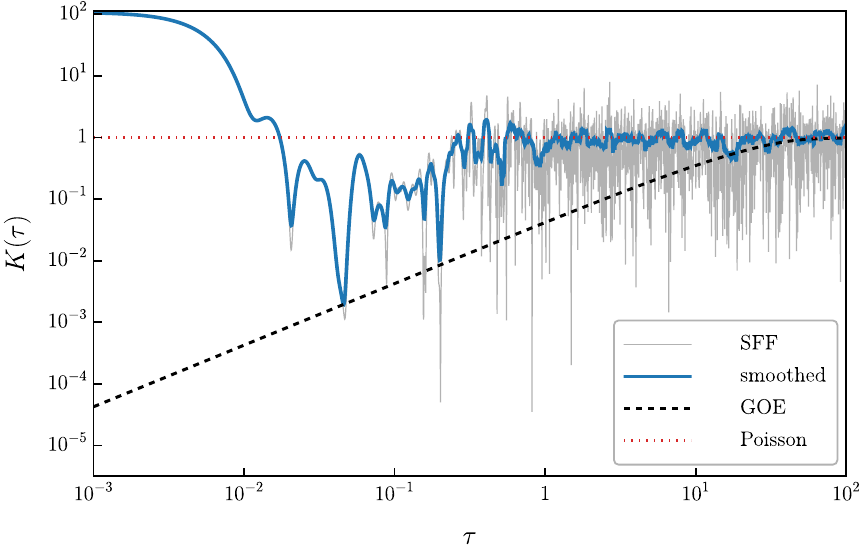}
            \caption{$(++)$}
        \end{subfigure}
        \par\smallskip
    \end{minipage}
    \hfill
    \begin{minipage}[t]{0.42\textwidth}
        \centering
        unfolded\par\medskip

        \begin{subfigure}{\linewidth}
            \includegraphics[width=\linewidth]{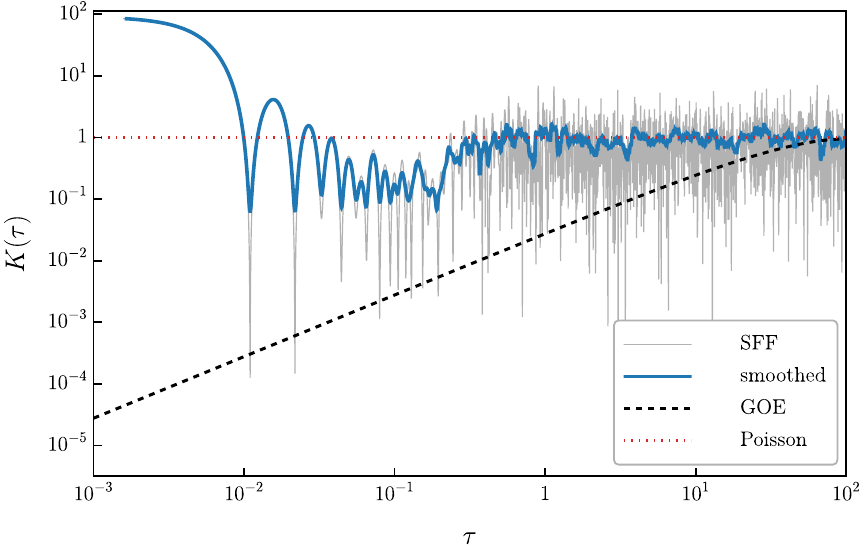}
            \caption{$(--)$}
        \end{subfigure}
        \par\smallskip

        \begin{subfigure}{\linewidth}
            \includegraphics[width=\linewidth]{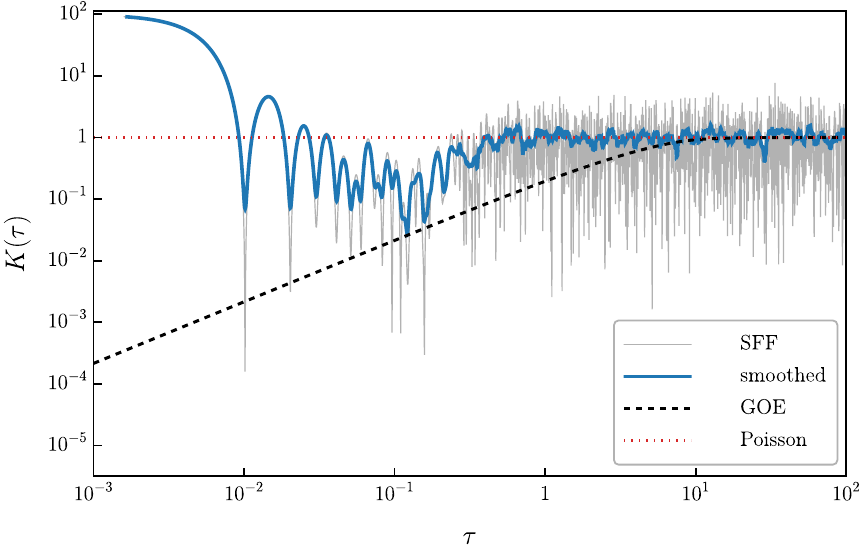}
            \caption{$(-+)$}
        \end{subfigure}
        \par\smallskip

        \begin{subfigure}{\linewidth}
            \includegraphics[width=\linewidth]{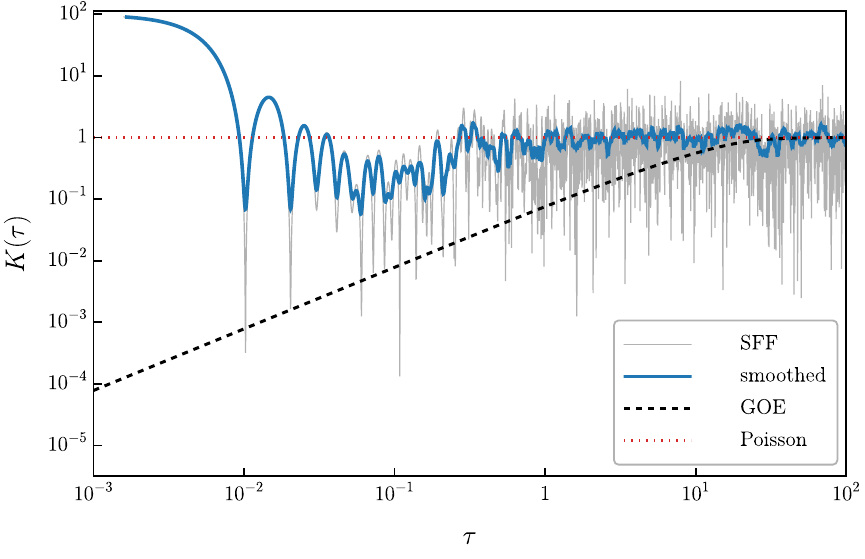}
            \caption{$(+-)$}
        \end{subfigure}
        \par\smallskip

        \begin{subfigure}{\linewidth}
            \includegraphics[width=\linewidth]{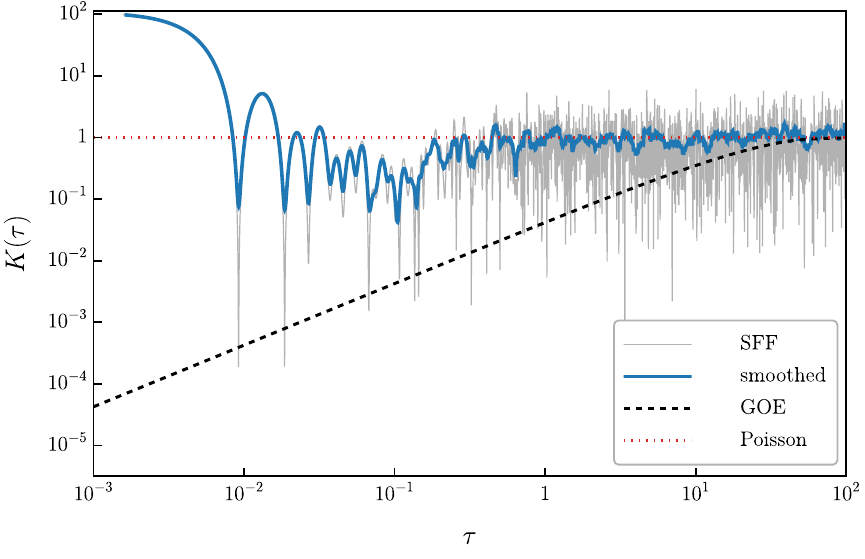}
            \caption{$(++)$}
        \end{subfigure}
        \par\smallskip
    \end{minipage}

    \caption{SFF $K(\tau)$ (Heisenberg time $\tau_H = 1$) in four symmetry sectors~\eqref{eq:H-ellipse}: $(--)$, $(-+)$, $(+-)$, $(++)$. The raw, i.e., physical SFF (left panels) clearly have no ramp in the $(+-)$ and $(++)$ cases, while the $(--)$ and $(-+)$
    cases are open to debate. The unfolded spectra (right) clearly
    show no ramp and thus no RMT-like correlations.}
    \label{fig:sector_sff}
\end{figure}

A typical SFF of the normal mode spectrum for the four symmetry sectors in the setup with an elliptical wall is shown in fig.~\ref{fig:sector_sff}. 
We solve eq.~\eqref{eq:KG-ellipse} using the Chebyshev–Gauss collocation method by introducing a radial coordinate $R(s,\phi)=r_w(\phi)+s[r_c-r_w(\phi)]$, $s\in[0,1]$, with $r_c=12$. This means that the computational domain is mapped into the space between the inner elliptical wall~\eqref{eq:ell-wall} and the outer circular boundary with radius~$r_c$ -- such finite approximation to AdS boundary is implemented for numerical purposes. 
We employ 35 Chebyshev–Gauss points and 40 uniformly spaced Fourier points, retaining 400 positive, approximately real normal-mode frequencies. The parameters are $r_h=\ell=1$, $\beta=0$, $a=1.15$, and $b=1.05$. Numerical unfolding is performed using a fifth-degree polynomial fit.

Even before unfolding (left panels), there is no convincing linear ramp. One could argue that the sectors $(--)$ and $(-+)$ come very close, however, the slope is not exactly linear and fluctuations are strong. In any case, the unfolded spectra (right panel) clearly have no ramp. This calculation strengthens our conclusion that the convincing linear ramp of the raw SFF in BTZ black hole with a circular brickwall~\eqref{eq:brickwall-domain} is an artifact of the overall shape of the spectrum and is not related to a hypothetical underlying RMT statistics.

\section{Normal modes of D1-D5 3-charge superstratum}\label{sec:ss}

The brickwall is meant to be a crude stand-in for the cap of a horizonless microstate geometry: a reflecting inner boundary at the bottom of a long but finite throat. Thus, our goal is to say as much as possible about actual fuzzballs, where instead of a brickwall we have an actual supergravity solution with a cap. Superstrata \cite{Bena:2016ypk,Bena:2017xbt} provide an excellent testing ground for the purpose.

We will now study the SFF of the normal modes of the superstrata. In this way, we can benchmark the results of the brickwall toy model and disentangle the influence of removing the horizon from the influence of actual microscopic dynamics. The superstrata are 3-charge $1/8$--BPS solutions in D1-D5 theory. In the supergravity regime, they are obtained as smooth horizonless solutions of 6-dimensional supergravity with the charges of the supersymmetric D1-D5-P black hole, and they have a long AdS${}_2$ throat that ends with a smooth cap. Their CFT duals are explicitly known states of D1-D5 CFT \cite{Shigemori:2020yuo}. There is no wall and no excised region: the reflecting ``boundary'' is the smooth closing-off of the geometry itself, so the boundary conditions for probes are fixed by regularity rather than chosen by hand.

The simplest superstratum solution is labeled by 3 quantum numbers $k$, $m$, $n$. We focus on the $(k,m,n)=(2,1,n)$ family, whose null geodesics are not integrable and whose massless wave equation does not separate \cite{Bena:2017upb}. This makes it closer to the elliptic brickwall than the circular brickwall, which is integrable. If the dip-ramp-plateau structure were a property of capped (i.e., smooth) throats rather than of the wall, this is where one would expect to see it.

We work in the decoupling limit, where the geometry is asymptotically AdS${}_3\times\mathbb{S}^3$. The 6-dimensional metric takes the standard BPS form:
\be \label{eq:BPSmet}
    ds^2_6 = - \frac{2}{\sqrt{\mathcal P}}(dv+\beta) \left[ du + \omega + \frac{1}{2}\mathcal F (dv+\beta) \right]+ \sqrt{\mathcal P}ds^2_4, \quad u,v = \frac{t \mp y}{\sqrt{2}},
\ee
over a flat base written in bipolar coordinates:
\be
    ds^2_4 = \Sigma \left( \frac{dr^2}{r^2+a^2} + d\theta^2 \right) + \left(r^2+a^2 \right) \sin^2 \theta d\varphi_1^2 + r^2 \cos^2 \theta d\varphi_2^2,
\ee
where $\Sigma = r^2 +a^2 \cos^2 \theta$. The coordinates $r$, $\theta$ parametrize the throat and the cap, and $u,v$ are lightcone coordinates obtained by mixing time $t$ and the compact direction along the circle $y$. For the $(2,1,n)$ family, the 1-forms $\beta,~\omega$ and the functions $\mathcal F,~\mathcal P$ depend on $(r,\theta)$ only. Their explicit form can be found in \cite{Bena:2017upb}:
\bea
    &&\mathcal P = \frac{Q_1 Q_5}{\Sigma^2} ~\Lambda^2, \quad \Lambda^2 = 1 - \frac{2(n+1)a^4 b^2}{2a^2+b^2} \frac{r^{2n}}{\left(r^2 +a^2 \right)^{n+2}}\sin^2 \theta \cos^2 \theta, \\
    &&\mathcal F = - \frac{b^2}{a^2}H_0(r) + \frac{(n+1)b^2}{2} \frac{r^{2n}}{(r^2+a^2)^{n+2}}\Sigma, \quad H_0 (r) = 1 - \frac{r^{2n+2}}{(r^2+a^2)^{n+1}}, \\
    && \omega_1 = \frac{R_y\sin^2 \theta}{\sqrt{2}\Sigma} \left [ a^2 +b^2 - \frac{b^2 r^{2n}}{(r^2+a^2)^{n+1}} \left( \frac{r^2}{2} + (n+1) a^2 \cos^2 \theta \right) \right ],\\
    && \omega_2 = \frac{R_y\cos^2 \theta}{\sqrt{2}\Sigma} \left [ a^2 + \frac{b^2 r^{2n+2}}{(r^2+a^2)^{n+2}} \left( \frac{r^2+a^2}{2} + (n+1) a^2 \sin^2 \theta \right) \right ], \\
    && \beta = \frac{R_ya^2}{\sqrt{2}\Sigma} \left( \sin^2 \theta d\varphi_1 - \cos^2\theta d\varphi_2\right).
\eea
Smoothness of the cap requires $Q_1Q_5/R_y^2 = a^2+b^2/2$, where $R_y$ is the radius of the $y$-circle and $b$ measures the momentum wave. We set $\sqrt{Q_1Q_5}= R_y = 1$, so that $b^2 = 2 \left(1-a^2 \right)$. The geometry then has three regions: an AdS${}_3\times\mathbb{S}^3$ region at large $r$, the AdS${}_2$ throat for $a \lesssim r$, and the cap at $r \lesssim a$. The throat depth is:
\be \label{eq:ll1}
    \frak{l}\equiv\frac{\sqrt{Q_1 Q_5}}{a R_y} = \frac{1}{a},
\ee
and the limit $a \to 0$ at fixed charges approaches the black hole geometry. The parameter $a$ plays a role roughly similar to the distance of the wall from the horizon in brickwall models, except that nothing is removed from the geometry.

\subsection{Wave equation}

The metric has four Killing vectors $\{\partial_u,\partial_v,\partial_{\varphi_1},\partial_{\varphi_2} \}.$ For a massless scalar probe we take:
\be
    \Phi = e^{i (p_u u+p_v v+ q_1 \varphi_1+q_2 \varphi_2)} f(r,\theta), \quad p_u =p_v = \omega/\sqrt{2}, \quad q_{1,2} \in \mathbb Z,
\ee
so that $\Phi \propto e^{i \omega t}$ carries no momentum along $y$. A property of every metric of the form~\eqref{eq:BPSmet} is that the components of $\sqrt{-g}g^{MN}$ coincide with those of the flat base~\cite{Bena:2017upb}. The wave equation therefore reduces exactly to:
\be \label{eq:KGss}
\frac{1}{r} \partial_r \left[ r \left(r^2+a^2\right) \partial_r f \right] + \frac{1}{\sin \theta \cos \theta} \partial_{\theta}\left( \sin \theta \cos \theta \partial_\theta f \right) + U(r,\theta)f = 0,
\ee
where $U=-\Lambda G^{AB}p_Ap_B$ and $G_{AB}$ is the $4\times 4$ block of the metric along the Killing directions and $p_A=(p_u,p_v,q_1,q_2)$\footnote{We solve~\eqref{eq:KGss} by Chebyshev-Gauss collocation method along both axes, introducing the transformed coordinates $x = r^2/(r^2+a^2) \in (0,1)$ and $w= \cos(2 \theta) \in (-1,1)$.}. The potential is quadratic in $\omega$:
\be \label{eq:pot1}
    U = - \frac{q_1^2}{\sin^2 \theta} - \frac{q_2^2}{\cos^2 \theta} + A_0(r) + \omega U_1(r,\theta)+\omega^2 A_2 (r).
\ee
All $\theta$-dependence beyond the $\mathbb{S}^3$ centrifugal terms sits in the term $\omega U_1$. This term is the obstruction to separability \cite{Bena:2017upb}: it is proportional to $\omega (q_1 +q_2)\cos (2 \theta)$. On the locus $q_1 = - q_2$ it vanishes and the problem separates into the round-$\mathbb{S}^3$ harmonics times radial towers.

Normalizable modes decay as $f \sim r^{-2-\ell}$ at the AdS${}_3$ boundary. Regular modes behave as $f \sim r^{|q_2|}$ at the cap and as $\sin^{|q_1|}\theta ~\cos^{|q_2|}\theta$ at the poles of $\mathbb{S}^3$. These are the smoothness conditions, and together with~\eqref{eq:KGss} they define a real discrete normal mode spectrum $\{\omega_i\}$.

The contrast with the brickwall is already visible at this stage. Fixing $(p_u-p_v,q_1,q_2)$ fixes every conserved charge, so each triplet labels one symmetry sector. Within each sector, the remaining 2-dimensional problem has no further symmetry, and its spectrum is a single ordered sequence. There is no need to pick a 1-dimensional subsequence such as $\mathcal S_{J_0}$ or $\mathcal S_{n_0}$ in (\ref{eq:spectral-subsequences}). In particular, the analysis below never mixes symmetry sectors.

The spectrum has two structural features that matter for the SFF. First, the smooth counting function grows as power law, $\bar N(\omega)\propto \omega^\alpha$ with $\alpha\simeq 2.0$, as expected for a two-dimensional cavity. It is not logarithmic as in (\ref{logspec}). Second, the lowest mode sits at $\omega_1 \simeq 2 a^2 = 2/l^2$, the familiar parametrically small gap of microstate geometries~\cite{Bena:2018bbd}. Above the gap, the spectrum is organized by the $\mathbb{S}^3$ harmonic $\ell$. In the separable sector $q_1 = -q_2$, each $\ell$ carries a single radial tower $\omega_{\ell,N} =a^2 \varepsilon_{\ell,N}+ \mathcal O(a^4)$, starting at $\varepsilon_{\ell,0} = \ell +2 + \delta_{\ell}$ with a small, decreasing shift $(\delta_2 \simeq 0.11,~\delta_{12} \simeq 0.007)$. The towers are not arithmetic: the spacing $\varepsilon_{\ell,N+1}- \varepsilon_{\ell,N}$ grows monotonically, from $\simeq 2$ at the bottom of the tower to an $\ell$-independent asymptote $d_{\infty}(n)$, approached as\footnote{For example: $d_\infty(n=1,2,5, 10) \simeq 3.2003,~2.8603,~2.5313,~2.364.$} $d_\infty - c_\ell/N^2$. At $b=0$ the tower reduces to the global AdS$_3$ ladder $\varepsilon = \ell + 2 + 2N$; the momentum wave stretches them, more strongly at high radial excitations.

\begin{figure}[h!]\centering
    \subfloat[raw]{\includegraphics[width=0.48\linewidth]{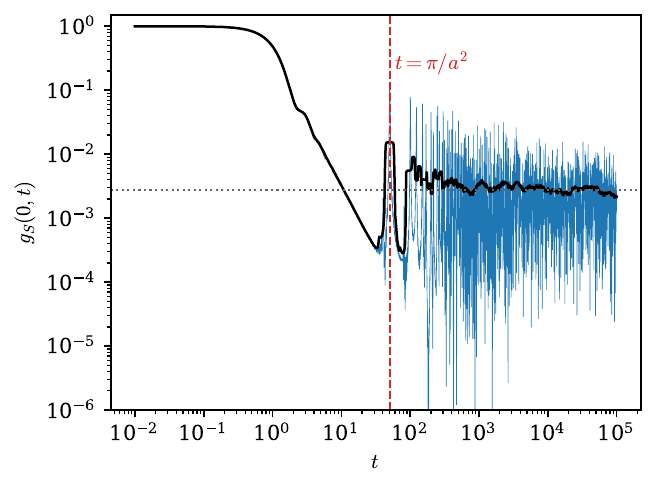}}\quad
    \subfloat[unfolded]{\includegraphics[width=0.48\linewidth]{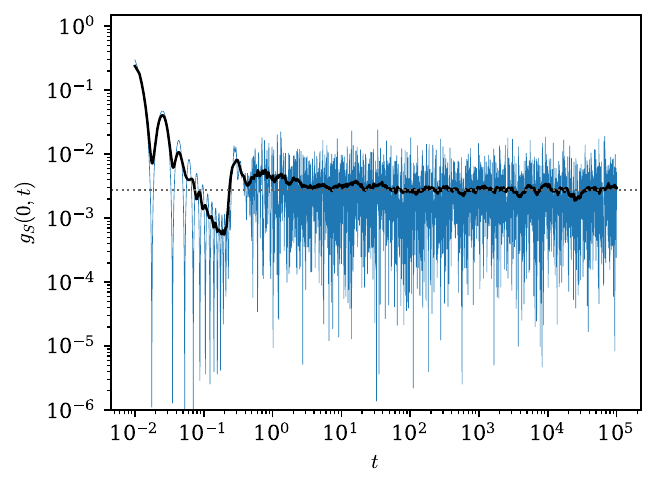}}
    \caption{SFF of the normal modes of the $(2,1,5)$ superstratum with ${a=0.25}$, $l=4$, in the sector $(q_1,q_2) = (1,0)$ with altogether $N_{\mathcal S}=357$ levels, computed from the raw, i.e., physical spectrum (a) and from the unfolded spectrum (b). Blue: single spectrum; black: moving average in $\log t$; dotted: plateau $1/N_{\mathcal S}$. In (a) the decay is followed by a jump to the plateau at the cap revival time $t = \pi/a^2$ (dashed), with no ramp in between. In (b) the levels are unfolded to unit mean spacing but the overall behavior is similar. Compare with fig.~\ref{fig:unfolded_WKB} for the brickwall.}
    \label{fig:ss1}
\end{figure}

\subsection{Spectral statistics}

Fig.~\ref{fig:ss1}(a) shows the SFF of the physical (raw) spectrum, computed exactly as in Section \ref{sec:circbrick}. After the initial decay, the moving average falls below the plateau and then jumps back to it abruptly at a time which can be estimated as:
\be \label{eq:sstrev}
    t_{\mathrm{rev}} \simeq \frac{2\pi}{2 a^2} = \pi \mathfrak l^2.
\ee
This is the period of the cap modes. For $r \ll a$ the functions $\Lambda \to 1$ and $\mathcal F  \simeq - b^2/a^2$, and the momentum wave terms in $\omega$ are suppressed by $(r/a)^{2n}$, and with $r = a \sinh \rho$, the~$(u,v,r)$--part of the metric becomes global AdS$_3$. Normal modes of global AdS$_3$ are spaced by $2$ in the cap time $T = t/\mathfrak l^2$, hence by $2/\mathfrak l^2$ in $t$. Therefore, $t_{\mathrm{rev}}$ is the light-crossing time of the cap: $\pi$ redshifted by $\mathfrak l^2$. The approach to the plateau is a revival, not a ramp. This is consistent with Section~\ref{sec:circbrick}: the brickwall ramp before unfolding comes from the logarithmic smooth density, and that ingredient is absent here, because the smooth cap produces a power-law density.

\begin{figure}[h!]\centering
    \subfloat[]{\includegraphics[width=0.48\linewidth]{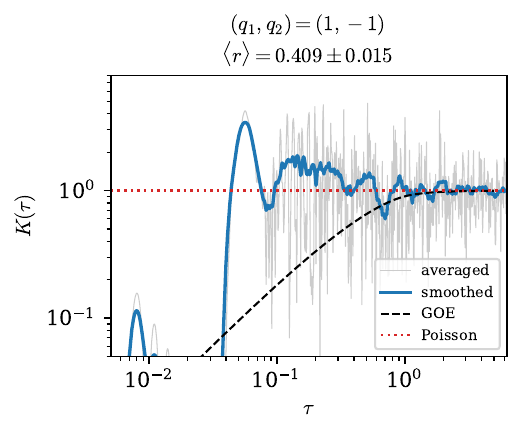}}\quad
    \subfloat[]{\includegraphics[width=0.48\linewidth]{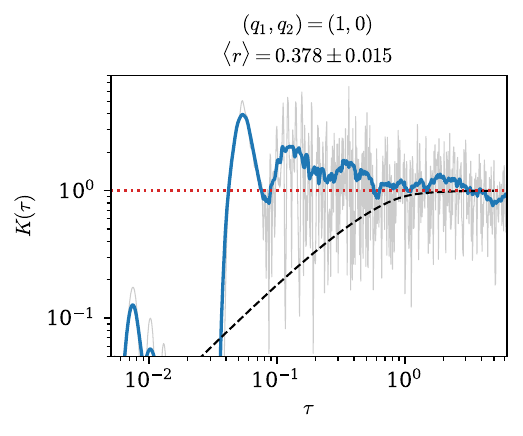}}\\
    \subfloat[]{\includegraphics[width=0.48\linewidth]{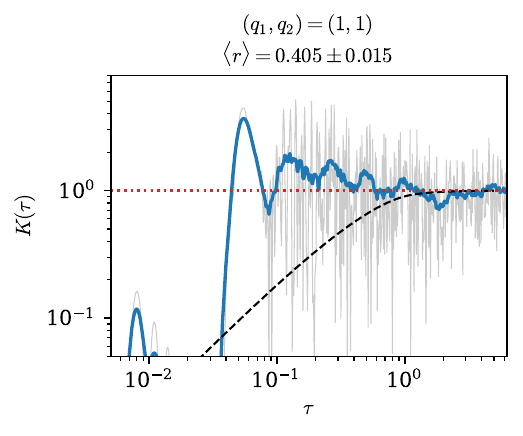}}\quad
    \subfloat[]{\includegraphics[width=0.48\linewidth]{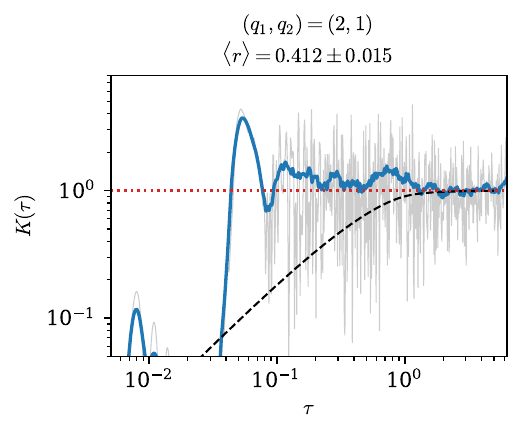}}
    \caption{The unfolded SFF $K(\tau)$ (Heisenberg time $\tau_H=1$) of the $(2,1,5)$ superstratum in the four symmetry sectors (a)-(d), averaged over seven backgrounds with $a \in [0.235,0.265]$ and zoomed-in to show the details of the revival epoch. Gray: averaged; blue: additionally smoothed in $\log \tau$. For comparison we plot the GOE ramp (dashed) and the Poisson plateau (dotted). The sector $(1,-1)$ (a) is exactly separable; the other three sectors are not.}
    \label{fig:ss2}
\end{figure}

Figs.~\ref{fig:ss1}(b) and \ref{fig:ss2} show the formal, unfolded spectrum. Fig.~\ref{fig:ss1}(b) is for a single realization (just like the panel (a) in the same figure). In fig.~\ref{fig:ss2} we average over seven backgrounds with different parameter $a \in [0.235,0.265]$. This produces an alternative to a moving average for obtaining a smoothed SFF. In all four sectors, including the three non-separable ones, $K(\tau)$ lies at or above the Poisson value for $\tau <1$, with no trace of the GOE ramp. The dominant feature is the clustering peak at $\tau \simeq 0.055$. Since $\tau$ is measured in units of Heisenberg time, this corresponds to a structure that repeats every $1/ \tau \simeq 18$ unfolded mean spacings, out of $N_\mathcal S \simeq 350$ levels in total, i.e., one period of the cap tower $\Delta \omega \simeq 2 a^2$ (eq.~(\ref{eq:sstrev})) contains about 18 levels. This is the cap tower spacing once again: the unfolded spectrum remembers the cap as level clustering \cite{BerryTabor1977}, not as level repulsion. This is further corroborated by the fact that the separable sector $(1,-1)$ shows the same behavior as the other, nonseparable sectors.

The clearest evidence of the absence of chaos in normal modes is the ratio $r_i = \mathrm{min}(s_i,s_{i+1})/\mathrm{max}(s_i,s_{i+1})$ of consecutive spacings \cite{Atas:2013gvn,Bianchi:2023uby,Ageev:2024gem}. Its mean is $0.386$ for Poisson and $0.531$ for GOE. Table \ref{tab:ss-r} collects our results as a function of throat depth and sector. All values are close to Poisson within error margins: in particular, there is no drift towards GOE as $l$ grows, and the nonseparable sector cannot be distinguished from the separable one.

\begin{table}[t]
\centering
\begin{tabular}{c c | c c}
\hline
$a$ & $\mathfrak l=1/a$ & $\langle r\rangle$, $(q_1,q_2)=(1,-1)$ & $\langle r\rangle$, $(q_1,q_2)=(1,0)$\\
\hline
$0.35$ & $2.86$ & $0.366\pm0.014$ & $0.401\pm0.014$\\
$0.25$ & $4.00$ & $0.401\pm0.015$ & $0.369\pm0.014$\\
$0.15$ & $6.67$ & $0.381\pm0.014$ & $0.395\pm0.014$\\
\hline
\multicolumn{2}{c|}{ensemble $a\in[0.235,0.265]$} & $0.409\pm0.015$ & $0.378\pm0.015$\\
\hline
\end{tabular}
\caption{Mean spacing ratio for the $(2,1,5)$ superstratum, with $\simeq330$--$380$ levels per entry. The ensemble values for $(q_1,q_2)=(1,1)$ and $(2,1)$ are
$0.405\pm0.006$ and $0.412\pm0.006$. Reference values: Poisson $0.386$, GOE $0.531$.}
\label{tab:ss-r}
\end{table}

\subsection{Momentum-fractionated superstratum: ramp with level repulsion}

\begin{figure}[h!]\centering
    \subfloat[]{\includegraphics[width=0.45\linewidth]{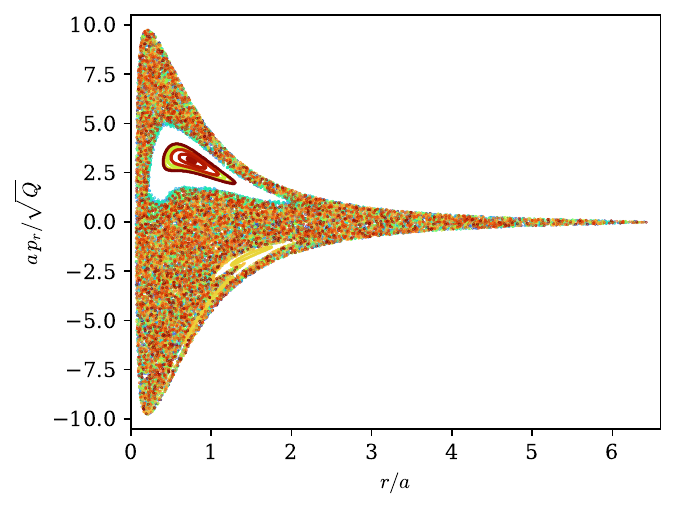}}\quad
    \subfloat[]{\includegraphics[width=0.45\linewidth]{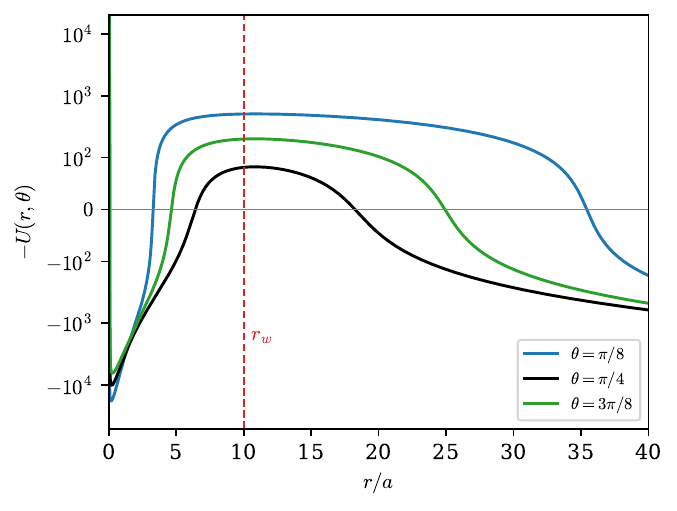}}
    \caption{(a) Poincar\'e section of null geodesic flow with parameters from eq. (\ref{eq:fsparams}), adapted from \cite{Djukic:2026phb}. (b) The effective potential of the same momentum-fractionated background, showing a deep potential well near the cap: the dynamics is dominated by the complicated and nonintegrable cap structure.}
    \label{fig:momfracpot}
\end{figure}

The $(2,1,n)$ single-mode family is the simplest superstratum, and its null geodesics are only weakly nonintegrable: the Poincar\'e sections are almost foliated by invariant curves, with only thin stochastic layers \cite{Djukic:2026ejk}. A sharper test of the claim that the dip-ramp-plateau structure is \emph{not} a property of capped throats in a horizonless geometry whose probe dynamics is unambiguously chaotic. Momentum-fractionated superstrata \cite{Bena:2016agb} provide one. These are smooth 1/8-BPS solutions in which the momentum wave is carried by $1/k$-fractionally moded excitations of a $k$-wound D1-D5 supertube. In the simplest member of the family (coiffuring Style 1, $Q_1=Q_2\equiv Q$) the metric is again of the BPS form (\ref{eq:BPSmet}), \mbox{cohomogeneity-2} in $(r,\theta)$, with four commuting Killing vectors, and it is specified by $(k,n)$, the cap scale $a$, and the momentum-wave amplitude $b_4$. The reduced equation (\ref{eq:KGss}) holds verbatim, with $U =- \sqrt{\mathcal P}\Sigma G^{AB}p_Ap_B$ now depending on $(r,\theta)$ in all three coefficients of its expansion in $\omega$. The explicit functions $\mathcal P,\mathcal F,\beta,\omega$ are listed in Sec.~2.2. of \cite{Djukic:2026phb}. For $kn \geq 2$ the null-geodesic Hamilton-Jacobi equation does not separate, and in the asymptotically flat solution at the values of the conserved momenta studied in \cite{Djukic:2026phb}, the Poincar\' e sections show a single connected chaotic sea with one surviving island (fig.~2(b) there, fig.~\ref{fig:momfracpot}(a) here). Motivated by general lessons from periodic orbit theory accounts on the necessary conditions for the presence of a ramp in SFF \cite{Richter:2001ojn,Haneder:2024ccc}, we take precisely this background and the same energy shell:
\be \label{eq:fsparams}
    kn=5,~\lambda \equiv \frac{2|b_4|}{Q}=0.9,~D\equiv \frac{Q}{a}=100,~\frac{m_1}{Q}=0.1,~\frac{m_2}{Q}=-0.067,
\ee
in units $a=1$ (so $Q = 100,~|b_4|=45$). Here $\lambda \to 1$ is the black hole scaling limit $Q_p \to Q_1 Q_2 /k^2$ at $a \to 0$, and $D$ is the depth of the AdS$_2$ throat, the analogue of $\mathfrak l$ in (\ref{eq:ll1}).

One subtlety is worth commenting. This solution is asymptotically flat rather than asymptotically AdS$_3$ (i.e., we have not taken the decoupling limit). At fixed conserved momenta the eikonal potential $U(r,\theta)$ of (\ref{eq:pot1}) (displayed on fig.~\ref{fig:momfracpot}(b)) is positive in the cap $(r \lesssim 6a)$, negative in a band $7a \lesssim r \lesssim 15a$ for all $\theta$, and positive again at infinity. Probe waves are therefore quasi-bound: they are trapped in the cap by a potential barrier and leak to infinity only by tunneling. The real parts of quasinormal frequencies are obtained to the accuracy needed here by imposing a Dirichlet condition at $r_w = 10a$. As before, we are interested in computing the normal mode spectrum on this geometry. Thus, the reflecting ``wall'' of the brickwall model is replaced by a smooth cap on the inside and by a genuine tunneling barrier on the outside. Again, nothing is excised or walled-off, and no boundary condition is imposed by hand on the trapped region. 

\begin{figure}[h!]\centering
    \subfloat[raw]{\includegraphics[width=0.48\linewidth]{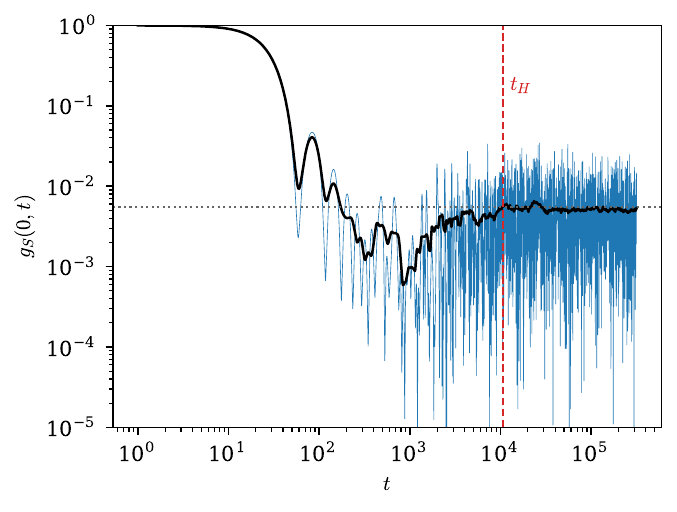}}\quad
    \subfloat[unfolded]{\includegraphics[width=0.48\linewidth]{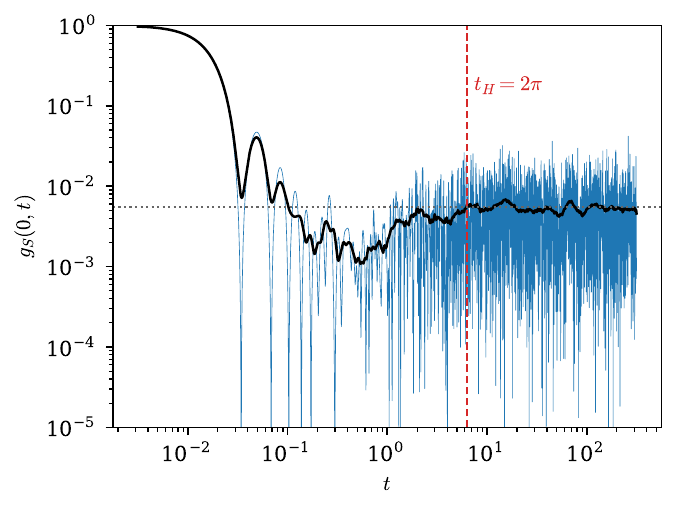}}
    \caption{SFF $g_{\mathcal S}(0,t)$ of the quasi-bound modes of the momentum-fractionated superstratum, sector $(q_1,q_2) = (6,-4)$ with altogether $N_{\mathcal S}=183$ levels, computed from the raw (a) and unfolded (b) spectrum. Blue: single spectrum; black: moving average in $\log t$; dotted: plateau $1/N_{\mathcal S}$; dashed: Heisenberg time. In both panels the moving average dips below the plateau and then rises back to it, reaching the plateau at $t \simeq t_H$.}
    \label{fig:momfrac1}
\end{figure}

Fig.~\ref{fig:momfrac1} shows $g_\mathcal S(0,t)$ for the raw spectrum. The initial decay and the oscillations reflect the finite spectral window and carry no information on correlations. But then we see something new compared to fig.~\ref{fig:ss1}: the moving average drops below the plateau and then climbs back \emph{gradually}, over roughly one decade, reaching $1/N_\mathcal S$ at $t \simeq t_H = 2\pi \bar \rho \simeq 1.1\times 10^4$ in units of $a$. For the $(2,1,n)$ superstratum, the same moving average also falls below the plateau, but then returns to it in a single step at the throat revival time. Here, in contrast, there is no sharp revival but a ramp, and its endpoint is the Heisenberg time, as it should be for a ramp originating from level repulsion. Fig.~\ref{fig:momfrac1}(b) confirms that this is not a consequence of the smooth overall shape of the level density: after unfolding to unit mean spacing, the same dip-ramp-plateau shape is present, with plateau reached at $t=2\pi$. For the brickwall, in contrast, the raw ramp disappears upon unfolding, because it is produced by the logarithmic smooth density.

\begin{figure}[h!]\centering
    \subfloat[]{\includegraphics[width=0.48\linewidth]{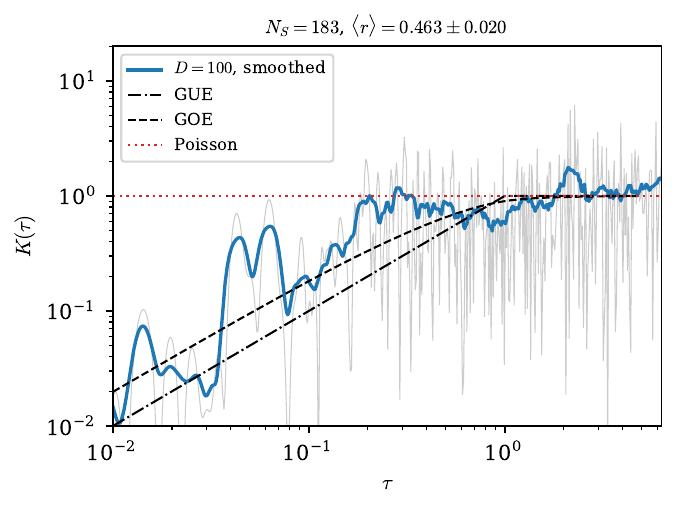}}\quad
    \subfloat[]{\includegraphics[width=0.48\linewidth]{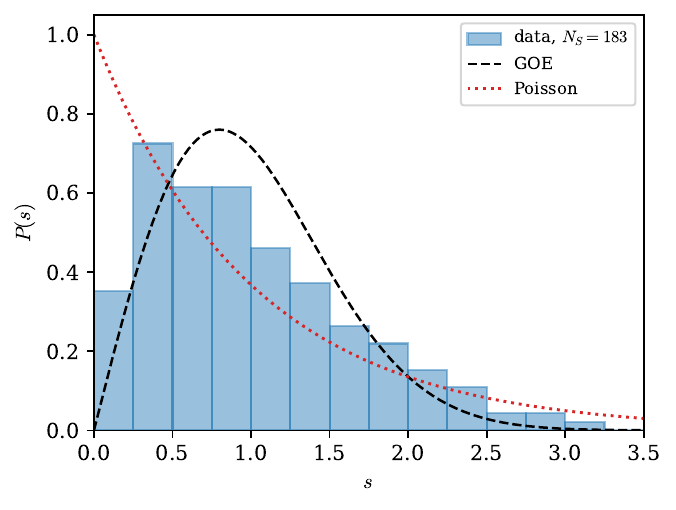}}
    \caption{(a) Unfolded SFF $K(\tau)$ (Heisenberg time $\tau=1$) of the momentum-fractionated superstratum with conventions of fig.~\ref{fig:momfrac1}. Gray: filtered SFF; blue: additionally smoothed in $\log \tau$; dashed-dotted: GUE, $K=\tau$; dashed: GOE, $K = 2 \tau - \tau \log (1 +2\tau)$; dotted: Poisson, $K=1$. (b) The level-spacing histogram follows the Wigner-Dyson shape at small $s$, a peak at $s \simeq 0.5$, and a tail that dies faster than Poisson, but is not GOE. This is expected for a system with mostly chaotic, but still mixed phase space in the semiclassical limit.}
    \label{fig:momfrac2}
\end{figure}

Fig.~\ref{fig:momfrac2} again shows the unfolded form factor, computed exactly as in fig.~\ref{fig:ss2}, but zoomed-in at the ramp. For $0.03 \lesssim \tau \lesssim 0.3$ the smooth $K(\tau)$ lies well below the Poisson value, and rises towards the plateau, which it reaches at $\tau \simeq 0.4-0.5$. Over the same range of $\tau$ every sector of the $(2,1,n)$ superstratum gives $K \geq 1$ (fig.~\ref{fig:ss2}). The curve lies above the GUE line everywhere and crosses the GOE curve from above. This has a simple origin in the classical phase space. The surface of section (fig.~\ref{fig:momfracpot}(a)) is not fully chaotic, namely a regular island survives around stable periodic orbits. A spectrum associated with a mixed phase space is, to leading order, a superposition of an uncorrelated (Poisson) sequence carrying the regular fraction of the levels and a random matrix sequence carrying the chaotic fraction~\cite{BerryRobnik}\footnote{We thank Juan-Diego Urbina for a discussion on this point.}. The former contributes a constant $\simeq 0.15$ to $K$ at all $\tau$, the latter the ramp. The data in fig.~\ref{fig:momfrac2} are consistent with this expectation, and the same superposition is visible in the eigenvalue statistics: the mean spacing ratio is $\langle r \rangle= 0.463 \pm 0.020$, i.e. above the Poisson value $0.386$ and below the GOE value $0.531$ (see also fig.~\ref{fig:momfrac2}(b)), where the $(2,1,n)$ superstratum has $0.37-0.41$ in every sector.

These computations establish the conclusion that when the probe dynamics on a horizonless microstate geometry is sufficiently chaotic, its probe spectrum develops a dip, a ramp and a plateau that survive unfolding and come with level repulsion, exactly the combination that the brickwall does not produce. Linear ramps have been obtained in gravity before, from Lorentzian wormhole saddles: the double cone of \cite{Saad:2018bqo} in JT gravity. There the ramp is a property of the averaged $\langle Z(\beta+it)Z(\beta-it)\rangle$, produced by a connected two-boundary geometry, and no discrete spectrum is exhibited. Here, by contrast, the ramp is read off from the spectrum of a single smooth solution, with no averaging. To our knowledge, this is the first instance in which the normal-mode spectrum of a single smooth horizonless supergravity solution, computed from the Lorentzian wave equation without any averaging, exhibits a dip-ramp-plateau structure that survives unfolding and is accompanied by level repulsion.

\subsection{Lessons}

Set against the brickwall, the two studied cases of superstrata sharpen the conclusion of the previous sections in several ways.
\begin{enumerate}
    \item \emph{The cap does what the wall does, without excision.} Both superstrata have a reflecting bottom, a discrete spectrum and a tunable throat, the features that the brickwall indeed mimics. The modes, however, live on the complete geometry, and their boundary conditions are fixed by smoothness. The $(2,1,n)$ has no ramp, and the ramp of the fractionated superstratum has a different origin from that of the wall (see the item 3 below). The brickwall ramp is therefore not inherited from the horizonless capped throat which it is supposed to model.
    \item \emph{Symmetry sectors are resolved by construction.} Every continuous symmetry is diagonalized by $(p_u-p_v,q_1,q_2)$, and each sector gives a single sequence. Thus, both the absence and the presence of a ramp are single-sector statements, which was one of the concerns raised about the sequence (\ref{eq:spectral-subsequences}) for the brickwall.
    \item \emph{Nonintegrability is not enough; a chaotic sea is}. The circular brickwall is integrable and nevertheless shows a ramp. The $(2,1,n)$ superstratum is nonintegrable, but only weakly so. The momentum-fractionated superstratum, whose Poincar\' e section is dominated by a connected chaotic sea \cite{Djukic:2026phb}, shows a dip, a ramp and a plateau that survive unfolding and come with level repulsion. For capped geometries, the ramp thus indeed comes from the chaos in the probe dynamics in the cap. 
    \item \emph{Unfolding separates the two.} For the circular brickwall the ramp is present before unfolding and absent after it. For the superstratum, when the ramp is present, it is present both before and after unfolding.
    \item \emph{The cap enters as a revival time in one case and as a Heisenberg time in the other.} For the $(2,1,n)$ superstratum the one sharp imprint of the geometry on the SFF is the revival at $t_{\mathrm{rev}}\simeq \pi \mathfrak l^2$, together with the associated clustering peak after unfolding. For a brickwall model, the analogous scale is the round trip time in the tortoise coordinate $\simeq\log (2r_h /\varepsilon)$. Both are reflection times of the cavity set by the redshift near the reflecting region. For the fractionated superstratum, no such revival is visible; the moving average returns to the plateau gradually and reaches it at $t \simeq t_H$, the Heisenberg time of the sector.
\end{enumerate}

Two caveats limit the scope of these statements. First, both single-mode and fractionated-mode superstrata are atypical microstates: coherent states built from a handful of strand types \cite{Shigemori:2020yuo,Bena:2016agb}. Second, probe normal modes are not the spectrum of the microstate Hamiltonian. Our results therefore say nothing about the microstate SFF. It is worth recalling that the \textit{eigenfunctions} of probes on the $(2,1,n)$ backgrounds display Berry's random-wave behavior \cite{Djukic:2026ejk}, while the eigenvalues remain Poissonian. Nonetheless, the findings of this section go hand in hand with what was observed in \cite{Djukic:2026phb}, namely that the random wave chaos of the $(2,1,n)$ solution is clearly weaker compared to the one on fractionated backgrounds. A similar distinction is observed here with spectral statistics.

\section{Discussion and conclusions}
\label{sec:conclusions}

We have established a sequence of results with two main endgoals. The first goal is to understand the meaning of level dynamics of normal modes on smooth horizonless geometries -- fuzzballs and brickwall models. The second goal is to understand what, if anything, the brickwall model can tell us about black holes or fuzzballs, i.e. if the drastic simplification of the brickwall geometries as opposed to fuzzballs is justified or leads to a proverbial ``spherical cow'' model.

The answer to the first question is that the normal mode spectrum, somewhat surprisingly, knows about the chaotic properties of the background provided the background is a true microstate, i.e., a solution of supergravity equations. The telltale sign of chaos in the SFF -- the ramp -- is absent in a weakly nonintegrable fuzzball and present in a strongly nonintegrable (momentum-fractionated) fuzzball. Therefore, even though the levels of the probe have no direct relation to the levels of the microstate, they still yield some information on the microstate.

The message regarding brickwall models is that all apparently chaotic properties of this model depend in detail on how the excising of the horizon is done, i.e., on the geometry of the wall (wall position, shape of the wall and the resulting effective potentials). Our verdict on brickwall setups is that they indeed reproduce the main properties of fuzzballs such as a throat of variable length or a discrete normal mode spectrum, but do not offer information about fuzzballs and the black hole vs. fuzzball problem. 


Finally, we want to discuss our findings in the light of a broader picture of time scales in D1-D5 microstates. In the bulk, the boundary-to-boundary Green function of $(1,0,n)$ superstrata shows black-hole-like exponential decay at early times, followed by echoes from the cap \cite{Bena:2019azk}. The echoes are evenly spaced with a period of order $\propto 1/a^{2}= \mathfrak l^2$ and correspond to poles that lie on the real axis and are nearly linear. This is the same physics as the revival in fig.~\ref{fig:ss1}(a): a redshifted global AdS$_3$ cap produces an almost arithmetic ladder, and an arithmetic ladder produces echoes, not a ramp. In \cite{Bena:2019azk}, it was anticipated that more general nonseparable superstrata geometries would distort the echoes by scattering energy into other harmonics. For the $(2,1,n)$ family, separability breaking is weak and localized near the momentum wave \cite{Djukic:2026phb}, and the revival survives essentially intact. This intuition is confirmed for fractionated superstrata backgrounds we considered here, i.e. echoes are replaced by a ramp. At the opposite end of the moduli space, at the free orbifold point where the CFT is integrable, the 2-point function of light probes in typical 2-charge states does display a dip, a ramp, and a plateau \cite{Balasubramanian:2016ids}. In that case, the ramp is logarithmic and is built from the broad distribution of strand lengths in a typical state.

\acknowledgments

The authors would like to thank Iosif Bena, Dimitrios Giataganas, Yixuan Li and Juan~Diego Urbina for inspiring discussions. A.~I.~Belokon, V.~V.~Pushkarev and L.~Umrikhin would like to thank the Institute of Physics Belgrade for hospitality during the preparation of the manuscript. M.~\v{C}ubrovi\'c acknowledges support by the Delta ITP consortium, a program of the Netherlands Organization for Scientific Research (NWO) funded by the Dutch Ministry of Education, Culture and Science (OCW). Work at the Institute of Physics is funded by the Ministry of Education, Science and Technological Development and by the Science Fund of the Republic of Serbia. The work on Sections 3-4 was supported by Russian Science Foundation Grant No.\,24-72-10061 [https://rscf.ru/project/24-72-10061/] and performed at Steklov Mathematical Institute of the Russian Academy of Sciences (M.~\v{C} and V.~V.~P.).

We acknowledge the use of Claude Opus 5.5 and Fable 5.1 (Anthropic) in developing the numerical code for Section \ref{sec:ss}; all results were verified by the authors.

\bibliographystyle{jhep}
\bibliography{Paper.bib}
\end{document}